# An Overview of Analysis Methods and Evaluation Results for Caching Strategies

Gerhard Hasslinger[a], Konstantinos Ntougias[b], Mahshid Okhovat[c], Frank Hasslinger[c] and Oliver Hohlfeld[d]

[a] *Deutsche Telekom, Darmstadt, Germany*
[b] *University of Cyprus, Cyprus*
[c] *Darmstadt Univ. of Tech., Germany*
[d] *University of Kassel, Germany*

*Abstract*

We survey analytical methods and evaluation results for the performance assessment of caching strategies. Knapsack solutions are derived, which provide static caching bounds for independent requests and general bounds for dynamic caching under arbitrary request pattern. We summarize Markov- and time-to-live-based solutions, which assume specific stochastic processes for capturing web request streams and timing. We compare the performance of caching strategies with different knowledge about the properties of data objects regarding a broad set of caching demands. The efficiency of web caching must regard benefits for network wide traffic load, energy consumption and quality-of-service aspects in a tradeoff with costs for updating and storage overheads.

## 1. Introduction

The main purpose of caches is to speed up access to data and to reduce transport costs by storing data in fast storage technologies and/or close to the requesting users. Caching strategies select the most relevant cache content within limited cache space in order to optimize caching benefit [95][102].

The work on caching systems and their analysis started in a first phase with applications for support of CPU and database systems. A set of basic strategies, such as Least Recently Used (LRU) [88], First-In-First-Out (FIFO) [72] and clock-based schemes [31] were proposed and evaluated via simulation and Markov methods, which assume an Independent Reference Model (IRM) for the requests [52][72]. Belady's algorithm [7][68][75][86][90][113] provided a general hit ratio bound for caches and data items of fixed size over 50 years ago, following a "farthest next request first" eviction principle with full knowledge of the future request sequence.

Those early caching approaches were restricted to records or pages of equal size, until web caches started a new wave of work in a broader scope of wide area networks [1][2][16][17][23][38][67][77][102][116]. The hit ratio is the main cache efficiency measure, but web caches are optimized regarding traffic, delay reduction and other network-wide quality-of-service aspects as caching goals. Beyond the scope of this overview, cooperative caching networks have been established in clouds, content delivery and information centric networks (CDN, ICN) [6][18][46][64][84][100][128]. Content caching at the edge is foreseen in 5G/6G infrastructure as a crucial component for ultra-low delay services [94][97][107].

In response to flexibility demands for network-wide optimization, GreedyDual [1][2][17][18][67][83] and other score-based methods [15][37][57][116] were proposed. They use a set of information per object, such as the size, popularity, expected costs and benefits for delivering an object from the cache. On that basis, scores are computed for the relevance of objects, where those with highest scores are preferred as cache content. Such meta-data is evaluation in further elaborated in machine learning (ML) algorithms [71][74][104][125].

While the analysis becomes more complex for flexible and advanced caching methods, bounds for optimum caching efficiency have been derived from Belady's bound to general min-cost flow and knapsack solutions, which fully cover score-based strategies [4][7][11][19][58][59][92][93][112][116].

Time-to-live (TTL) caching emerged as another branch of strategies for web applications, where each object in a cache is valid only for a limited time. Pure TTL caches have varying size for storing all currently valid objects. Caches for the domain name service (DNS) and some other web applications with small and frequently updated data records are usually managed based on TTL, when sufficient storage space can be provided [6][20][26][76][108]. TTL analysis approaches are scalable even for hierarchical caching networks [6][12][20][27][84][101], because their hit and value ratio can be analyzed separately for each data object.

Figure 1 illustrates the relationship between caching strategies, applications and analysis schemes as the scope and the background structure of this work. For many omitted details we refer to caching policy surveys [95][102] and Table 1.

The aim of this overview is to present the current scope of performance analysis results for a single cache in the literature. Our main focus is on the progress made by the increasing work activity in this field over the last decade, which is apparent from Figure 2, yielding substantial extensions of classical solutions. Such progress can be noticed in

- recently developed min-cost flow [11] and knapsack bounds [59], which overcome unit data size restrictions of Belady's bound [7], and thus apply to usual web cache scenarios with varying data size, and with a value or utility being assigned as a cost/benefit measure per cacheable data item,

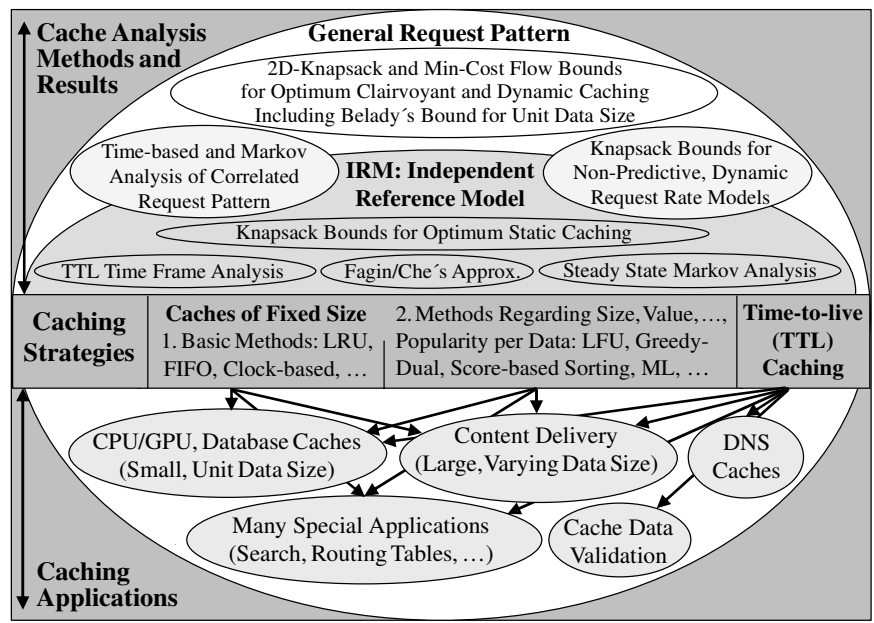

Figure 1: Scope of analysis methods for caching strategies (LFU: Least Frequently Used; GPU: Graphics Processing Unit)

- product form solutions for multi-level caches [51][78] extending IRM Markov analysis results for FIFO and RANDOM [52][72], which also hold for clock-based methods,

- and work on TTL caching analysis, most of which appeared in the last decade, with many cross relations to strategies for caches of fixed size, from LRU and FIFO hit ratio approximations [23][32][43] to general TTL-based utility concepts, corresponding to score-based methods with scores being mapped into TTL values.

The presentation of cache analysis methods in the main part is subdivided into the three main work streams according to those published results, see Figure 2 for a detailed time line:

(1) *General Bounds on the Cache Hit and Value Ratio*

We start from knapsack bounds for optimum static caches for IRM requests [4][92][93][99][116]. They are extended at first to varying request rates over time, and finally to arbitrary request pattern in 2-dimensional (2D-)knapsack solutions [19][59]. We compare the 2D-knapsack bound to similar results by Berger et al. [11] via minimum cost flow optimization. The bounds even include score- and utility-based caching goals and methods [34][93].

(2) *Markov Modeling and Analysis Results*

In a second part, steady state Markov analysis of caching performance is addressed. Results are derived for basic caching methods [32][52][72], for multi-level caches and networks, including cache startup, mixing and convergence times [8][12][50][78][105][110][115].

Cache filling phases are generally characterized by the Markov analysis for LRU [60], which is extendible to data of different size. Markovian product form solutions are scalable, but the complexity of other steady state results becomes intractable for usual and large cache sizes. While the derivation of basic Markov caching analysis is spread over several papers [52][72][73] in broad frameworks, we briefly present main results with their core proofs via equilibrium equations.

(3) *Time-to-live Caching Analysis*

A variety of TTL caching approaches and their analysis open new perspectives under different timing conditions and for various goals. TTLs were introduced by web servers to validate that data in caches can deviate from original data only for a limited time. TTLs can be adapted for controlling the amount of valid data in the cache or for an intended hit ratio and other purposes, such that TTL cache analysis is flexibly applicable in many ways [23][27][34][45][49][51][66].

In general, the flexibility to optimize caching benefits is important for many caching scenarios, while overheads also must be taken into account in terms of storage and upload capacities and the update effort of the applied caching strategy.

Figure 2 shows the intensity of work on the analysis of caching strategies in the literature over time. A first phase from 1965-1990 addressed basic caching methods for local CPU and database support. Then, web cache applications boosted work on new strategies from the mid 1990-ies with a broader scope on network wide cost and benefit analysis. Figure 2 indicates high activity of caching analysis studies especially in the last 5 years, although this picture is still incomplete. Some papers are listed twice in different categories of Figure 2, when they contribute to several areas.

There are a number of surveys and overview papers on basic caching strategies in the literature [38][95][101][102], but their focus is not on analysis methods and they have minor overlap with the main workstreams in this overview.

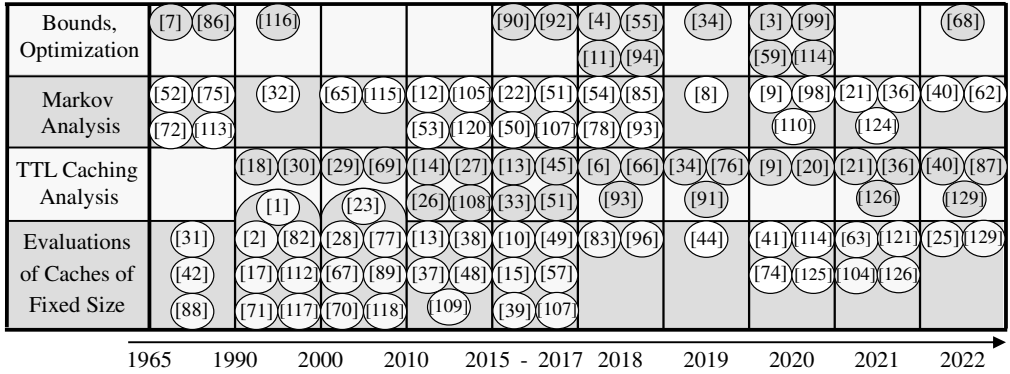

Figure 2: Time line of work on caching analysis topics in the literature

The analysis of caching networks in ICN and clouds is addressed in many surveys [23][35][64][80][97][100][128]. They investigate optimized placement of caches and content distribution in cache hierarchies, request routing and other network optimization aspects. Caching networks are not in our focus, but their performance analysis often builds on and extends solutions for single caches included in this overview, where each single cache can use network specific measures as input for optimization goals of score-/utility-based strategies.

We start the main part in Section 2 with demands for efficient caching. Knapsack bounds of the cache hit and value ratio are derived in Section 3 and extended to arbitrary request pattern in Section 4. Section 5 presents Markov analysis results of caching strategies in steady state and for convergence and mixing times, as well as extensions to multi-level and probabilistic caching. Section 6 addresses the impact of time-to-live conditions on caching performance and summarizes TTL cache analysis methods. Section 7 and the Conclusions compare different types of caching performance results, from which limitations and gaps in the state of the art are identified as prospective future research topics.

2. REQUIREMENTS FOR EFFICIENT CACHING STRATEGIES

Before considering specific analysis approaches, we address the requirements for efficient caching strategies as a starting point for relevant goals and performance measures. Caching methods should comply to the following demands, which are discussed in caching surveys and extended evaluation studies [2][10][17][37][38][57][89][95][102][112][116]:

(1) *Maximum Hit Ratio*
The cache hit ratio should be close to the optimum for the considered environment. This can be verified in comparison to the upper bounds provided in Sections 3-4. The simple bound of Eq. (1) and general knapsack bounds seem most relevant in practice [2][16][38][101][119].

(2) *Adaptability to Follow Dynamic Request Trends*
Caching methods have to react to changing content popularity. In this way, efficiency is extended beyond IRM towards correlated request pattern, where recent requests are more important than older ones. Even if IRM and static content may suit on medium time scales of hours up to one day [56][119], web caches must adapt to long-term churn in the working set of relevant objects.

(3) *High Update Speed and Low Overhead*
The implementation of a caching method should have a low update effort per request. In some cases, FIFO may even be preferred to LRU as the faster scheme despite of lower hit ratio [41][122]. CDNs supporting popular platforms like Wikipedia, YouTube or Akamai have to cope with workloads of millions of requests per second [6][58][130]. The meta-data overhead for cache management should be as small as possible.

(4) *Low Traffic Volume for Web Content Uploads*
Uploads of data with low caching benefit (low/zero-valued cache objects, one-timers or one hit wonders) should be avoided in web caches by proper content selection in order to reduce upload traffic and processing.

(5) *Flexibility to Adapt to Cost, Benefit and QoS Criteria*
Web caching methods should support performance optimization for network and quality of service (QoS) demands beyond the hit count. Therefore, the strategies should take the size, popularity, overheads, costs and benefits into account for serving objects from the cache.

The relevance of those requirements differs for the underlying applications. Local caches for support of CPU/GPU and database processing often refer to data units of fixed size and have special request pattern, e.g. with periodic repetitions. Strict demands for fast updates and low overhead are crucial for devices with limited power, and for caching in search and routing tables [106][108]. Web content delivery for downloads and streaming services usually can rely on powerful data center machines for high workloads of random Zipf distributed requests with data units (chunks, files) of largely varying size, where the demands (4) and (5) are important.

3. UPPER BOUNDS ON THE CACHE HIT AND VALUE RATIO FOR IRM AND VARYING OBJECT POPULARITY

Simple upper bounds of the cache hit ratio are obtained for independent requests via static caching, which can be extended to a changing set of objects with varying popularity. Bounds for optimum clairvoyant caching, which apply to arbitrary request pattern, are addressed in Section 4.

## 3.1. Maximum IRM Cache Hit Ratio for Unit Size Objects

The independent reference model is specified by the probabilities $p_k$ for requests to each object $O_k$ ($k = 1, …, N$) of a fixed object catalogue. $p_k$ is valid for the next and further request, independent of the past. The IRM hit probability is given by the sum of request probabilities $p_k$ over all objects, which are currently in the cache. When objects are sorted such that $p_1 \geq p_2 \geq … \geq p_N$ then the maximum hit ratio $h_{IRM}^{MAX}$ for a cache of size $M$ is obtained by storing the most popular objects $O_1, O_2, …, O_M$ in a static cache, i.e., without changing the cache content over time [52][82][101][117]:

$$h_{IRM}^{MAX} = p_1 + p_2 + … + p_M . \quad (1)$$

The least frequently used (LFU) cache eviction policy and score-based strategies [17][57][67] with the request count as the score are approaching the maximum of Eq. (1) in the long-term steady state behavior. Even in cases of weak correlation among requests, LFU is recommended on time scales up to one day [2][45][89][119].

Next, we compare the cache hit ratio bound of Eq. (1) to the performance of basic caching methods. The hit ratios of LFU, LRU, FIFO, and RANDOM caching methods are obtained in Figure 3 from long-term simulations over $10^8$ requests in each case. The evaluation excludes 10% of the request sequence from the start to avoid an impact of cache filling phases. In those simulations, we generate independent and Zipf distributed requests [16] according to

$$p_k = \alpha k^{-\beta} \quad (k = 1, …, N) \quad \text{with} \quad \alpha = 1/\Sigma_k k^{-\beta}, \quad (2)$$

which have often been confirmed for access to large web content platforms. Figure 3 includes three sets of curves for Zipf shaping parameters $\beta \in \{0.6, 0.8, 1\}$, covering the usual range for web applications [16][56].

The entire scale of the hit ratio curve (HRC) is shown over cache sizes from $10 \leq M < N = 10^6$. The steady state IRM hit ratio for FIFO and RANDOM caching have been proven to coincide via Markov analysis [52]. LRU hit ratios improve the FIFO results by up to 5%, as is generally proven for IRM [122]. LFU hit ratios are close to the upper bound of Eq. (1) and improve LRU by up to 16%. Small LRU hit ratios below 10% are more than doubled by applying LFU [121].

Evaluations for synthetic request pattern as in Figure 3, and for web request traces are usually performed separately for each cache size. Faster computation of the resulting hit or miss ratio curve (HRC, MRC) is suggested in [123][127] to enable online estimation of the working set and caching efficiency. LRU and LFU caching strategies allow to compute the hit ratio for arbitrary cache sizes in one step, because the cache content is obtained from the same stack or ordered list for any cache size. Hence, a request is a hit, if the position of the requested object is smaller or equal to the cache size in case of unit object sizes. This HRC evaluation approach is extendible to different object sizes and can be further speeded up by sampling techniques [123][127]. Then LRU/LFU cache hit ratios are obtained from counters for all considered cache sizes in parallel. The approach fails for many other caching methods, e.g. FIFO, because the top objects in a FIFO stack during a request trace depend on the cache size.

Moreover, an approximation formula of the cache miss ratio curve is proposed by [118], which assumes a specific dependency between the number of misses for a request workload and the cache size, and thus a specific format of the MRC. The average RAM- and disk-resident time, average time between reads of new pages and the average time (instructions executed) between a page's first entry and last exit from RAM are the parameters for computing the miss ratio approximation, which is applied and shown to fit well for RAM caching of CPU processing workloads, whereas suboptimal fit is reported for web retrieval workloads [118].

## 3.2. Web Caches for Data Objects of Different Size and Value

Next, we consider different sizes $s_1, …, s_N$ and specific values $v_1, …, v_N$ associated with the delivery of the objects $O_1, …, O_N$ from the cache, still for IRM requests with probabilities $p_1, …, p_N$. The values reflect costs and benefits of storing an object and serving it from the cache. When low delay is the main caching goal [97][112], the value $v_k^{Delay}$ should account for the delay difference between the original source ($d_{Source}$) and the cache ($d_{Cache}$), as well as an additional delay ($d_{Check}$) for checking whether an object is in the cache

$$v_k^{Delay} = [d_{Source}(O_k) - d_{Cache}(O_k)]|_{O_k \in C} - d_{Check}(O_k). \quad (3)$$

Recent studies on caching with delayed hits [3][33][40] due to latencies for fetching data from a server after a cache miss also express the caching value in terms of optimized delays, see Section 4.6 and the final part of Section 6.7 for more details. Traffic engineering can be considered as another goal, where the value $v_k^{TE}$ refers to the object size $s_k$ for measuring savings in the traffic volume and bandwidth on links and routers. Cache servers are crucial for load reduction especially on expensive international links in the Internet backbone. The object value can combine contributions for delay, traffic load, etc.: $v_k = v_k^{Delay} + v_k^{TE} + …$. Moreover, the optimization of a single cache with regard to network load and QoS demands is often integrated as a component in distributed caching infrastructures of CDNs, clouds and ICN architectures [35][64][79][80][84][97][101][128][130].

For optimizing cache value or utility goals, a corresponding performance measure is required. Let $r_1, …, r_R$ denote a request sequence of length $R$, such that $O_{rj}$ is the object referenced by the $j^{th}$ request with value $v_{rj}$ ($j = 1, …, R$). Then $V_C = \Sigma_{j: O_{rj} \in C} \, v_{rj}$ represents the value [59] or utility [93] due to serving objects from the cache $C$, which depends on the applied strategy. The total value corresponding to serving all requests from an infinite cache is $V_{Total} = \Sigma_j v_{rj}$. Then we de-fine the value hit ratio as $VHR = V_C/V_{Total}$. $VHR$ includes the object hit ratio $OHR$ as a special case for unit values $v_k = 1$, and the byte hit ratio $BHR$ for $v_k = s_k$. $OHR$ and $BHR$ represent the fractions of objects and bytes, respectively, being served from the cache during a request sequence. For IRM request pattern, $VHR$ can be expressed via $p_k$ and $v_k$ [58][93]

$$VHR_{IRM} = \sum_{\substack{k=1 \\ O_k \in C}}^{N} p_k v_k \Big/ \sum_{k=1}^{N} p_k v_k; \quad (4)$$

$$OHR_{IRM} = \sum_{\substack{k=1 \\ O_k \in C}}^{N} p_k; \quad BHR_{IRM} = \sum_{\substack{k=1 \\ O_k \in C}}^{N} p_k s_k \Big/ \sum_{k=1}^{N} p_k s_k.$$

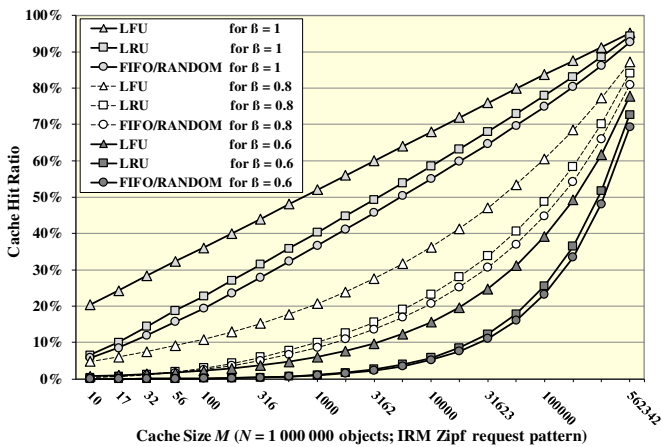

Figure 3: IRM hit ratio simulation results for unit object size

The results of Eq. (4) hold per request and as mean values for the entire request sequence. When $O_1, \ldots, O_N$ are sorted due to the scores $S_k = v_k p_k / s_k$ for expressing value density, such that $v_1 p_1 / s_1 \geq \ldots \geq v_N p_N / s_N$, then the maximum IRM value hit ratio $VHR_{IRM}^{MAX}$ is achieved by static caching of objects with the highest scores. We obtain [34][59][82][93][116]

$$\frac{\sum_{k=1}^{L} p_k v_k}{\sum_{k=1}^{N} p_k v_k} \leq VHR_{IRM}^{MAX} \leq \frac{\sum_{k=1}^{L} p_k v_k + q p_{L+1} v_{L+1}}{\sum_{k=1}^{N} p_k v_k} \leq \frac{\sum_{k=1}^{L+1} p_k v_k}{\sum_{k=1}^{N} p_k v_k} \quad (5)$$

where $s_1 + s_2 + \cdots + s_L \leq M < s_1 + s_2 + \cdots + s_{L+1}$
and $q = (M - s_1 + s_2 + \cdots + s_L)/s_{L+1} < 1$.

Eq. (5) represents a standard knapsack solution for filling a static cache of size $M$ with a maximum number of objects with highest value densities $v_k p_k / s_k$. The bound on $VHR_{IRM}^{MAX}$ meets the optimum value hit ratio, if $L$ top scored objects exactly fit into the cache ($\Rightarrow q = 0$).

Knapsack solutions for IRM request pattern are addressed manifold [4][34][59][82][92][93][99][116]. Tatarinov [116] suggests a knapsack algorithm with documents $D$ in the cache being sorted due to "*value(D) = benefit(D)/size(D)*" or in our notation: $S_D = v_D / s_D$. He estimates the knapsack method to "*maintain approximately optimal total cache benefit*", but as "*not efficient because it requires sorting of a large set of documents*". Nonetheless, GreedyDual caching methods were proposed to enforce exact sorting due to scores, whereas score-gated schemes [58] deploy a fast, approximate sorting method. Utility optimization approaches achieve approximate sorting with score-based preferences being enforced via TTL timers [34] or probabilistic insertion rules [93].

Figure 4 compares the knapsack bound of Eq. (5) with results for different caching strategies, again for independent Zipf distributed requests with $\beta = 0.75$. Objects of different size $s_k$ and unit value $v_k = 1$ are considered. We assume a log-normal distribution for object sizes with a mean of 622 kByte

$$s_k = e^{\mu + \sigma Z_k} \text{ kByte} \quad \text{with } \mu = 3.5, \sigma = 2.5, \quad (6)$$

where $Z_k$ is a standard normal distributed random variable. The size distribution is adapted to measurement statistics of cacheable CDN data chunks [10] and very similar results for web files reported in [74][112][116][125][130], which vary over a broad range from kByte to MByte. The popularity $p_k$ and the size $s_k$ of each object are assumed to be independent.

Figure 4 shows large hit ratio gaps between the knapsack bound and FIFO/RANDOM (up to 45%), LRU (up to 40%), and LFU (up to 30%). LFU prefers objects with highest request counts $c_k$, whereas a preference due to the score $c_k/s_k$ by a GreedyDual or score-gated clock strategy significantly improves the hit ratio towards the upper bound within < 5%.

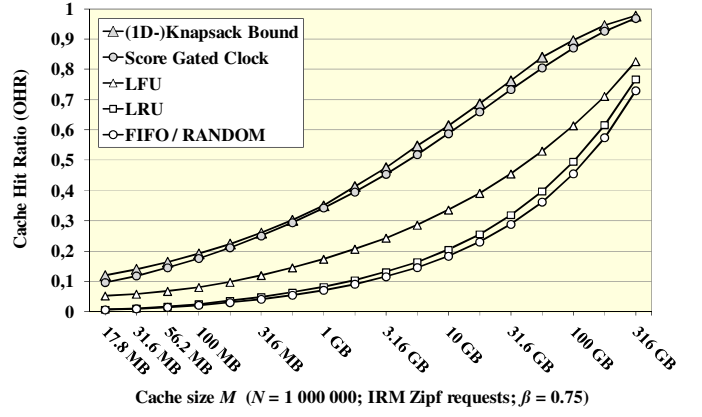

Figure 4: Simulated IRM hit ratios for objects of different size

### 3.3. Upper Bound on Caching Performance for Varying Object Popularity and Value over Time

The fact that static caching can be optimal beyond IRM assumptions is confirmed by time-based request models, which assume the requests per object to follow independent stationary point processes [45][99]. When the inter-request time distributions fulfill conditions of an increasing hazard rate function, then static caching is still proven to be optimal, see Theorem 1 in [45]. Hazard rate based bounds are derived in [99] for independent stationary point processes of the requests per object, yielding knapsack bounds for the byte and object hit ratio similar to the result of Eq. (4), see Section 3 in [99].

The previous bounds for independent requests can be straightforwardly extended with regard to varying request probabilities and object values over time. Let $p_1^{(r)}, \ldots, p_{N^{(r)}}^{(r)}$ and $v_1^{(r)}, \ldots, v_{N^{(r)}}^{(r)}$ denote the request probabilities and object values of the $r^{th}$ request ($r = 1, \ldots, R$). We include requests to new objects with probability $p_{New}^{(r)}$, such that $p_1^{(r)} + \cdots + p_{N^{(r)}}^{(r)} + p_{New}^{(r)} = 1$. The number of objects $N^{(r)}$ is incremented after each request to a new object, i.e., $N^{(r+1)} = N^{(r)} + 1$ with probability $p_{New}^{(r)}$, or otherwise $N^{(r+1)} = N^{(r)}$, with initial catalogue size $N^{(1)}$.

Under these assumptions, the knapsack bound of Eq. (5) still applies for each request, while the object ranking due to scores $S_k^{(r)} = v_k^{(r)} p_k^{(r)} / s_k$ may change per request due to changes in the components $v_k^{(r)}$ and $p_k^{(r)}$. Let $Rank_1^{(r)}, \ldots, Rank_{N^{(r)}}^{(r)}$ denote the object ranking sorted due to scores $S_k^{(r)}$ for the $r^{th}$ request: $S_{Rank_1^{(r)}}^{(r)} \geq S_{Rank_2^{(r)}}^{(r)} \geq \ldots \geq S_{Rank_{N^{(r)}}^{(r)}}^{(r)}$. Then we obtain

$$VHR^{MAX} \leq \frac{\sum_{r=1}^{R}(1-p_{New}^{(r)})\left(\sum_{j=1}^{L}p_{Rank_j^{(r)}}^{(r)}v_{Rank_j^{(r)}}^{(r)} + q^{(r)}p_{Rank_{L+1}^{(r)}}^{(r)}v_{Rank_{L+1}^{(r)}}^{(r)}\right)}{\sum_{r=1}^{R}\sum_{j=1}^{N^{(r)}}p_{Rank_j^{(r)}}^{(r)}v_{Rank_j^{(r)}}^{(r)}}$$

where $s_{Rank_1^{(r)}} + \cdots + s_{Rank_L^{(r)}} \leq M < s_{Rank_1^{(r)}} + \cdots + s_{Rank_{L+1}^{(r)}}$ (7)
and $q^{(r)} = (M - s_{Rank_1^{(r)}} - s_{Rank_2^{(r)}} - \cdots - s_{Rank_L^{(r)}})/s_{Rank_{L+1}^{(r)}} < 1$.

Requests to new objects are cache misses, which reduce the cache hit ratio by the factor $1 - p_{New}^{(r)}$. The format of Eq. (7) covers the content churn model with constant popularity used in [39] as well as the Markov popularity model in [107].

While the bound of Eq. (5) is always approached by static caching of $L$ top-scored objects, the extension of Eq. (7) is reachable only if the cache content can follow a changing set of top-scored objects $TopObj_L^{(r)} = \{O_{Rank_1^{(r)}}, \ldots, O_{Rank_L^{(r)}}\}$ in each request $r$. The changes can be followed if $\forall r \in \{1, \ldots, R-1\}$: $TopObj_L^{(r+1)} \subseteq TopObj_L^{(r)} \cup \{O^{(r)}\}$, where the $r^{th}$ request refers to $O^{(r)}$. An optimum strategy has to maximize the sum of scores of cached objects, i.e., an external object $O^{(r)}$ is inserted if and only if $O^{(r)}$ has higher score than the eviction candidates. The bound of Eq. (7) is flexible to model variable request pattern as well as new emerging objects, where dynamic object popularity can be specified by the probabilities $p_1^{(r)}, \ldots, p_{N^{(r)}}^{(r)}$.

Studies on web request pattern [47][96][119] confirm object popularity profiles characterized by a fast ramp up to a maximum, followed by a slow, long-lasting decrease. For modeling of such behavior, we can set $p_k^{(r+1)} = p_k^{(r)}(1 - p_{N^{(r+1)}}^{(r+1)})$ for $k = 1, \ldots, N^{(r)}$, if a new object enters at the $r^{th}$ request with initial request rate $p_{N^{(r+1)}}^{(r+1)}$. Otherwise, request rates remain unchanged $p_k^{(r+1)} = p_k^{(r)}$. Then the renewal rate $p_{New}^{(r)}$ and the initial request probabilities $p_{N^{(r+1)}}^{(r+1)}$ for new objects determine the intensity of churn in object popularity, while the popularity of old objects is fading away in favour of new ones. $p_{N^{(r+1)}}^{(r+1)}$ can be randomly chosen from a Zipf-like distribution for usual web request pattern [16].

Instead of a start for new objects at the maximum request rate, the bound of Eq. (7) is also adaptive to more general rate profiles [47][96][119] via Markovian rate models, similar to shot noise models. Time based request rate modeling [45][99] is addressed in more detail in Section 6 on TTL caching.

*3.4. Score-based Caching, GreedyDual or Utility Optimization*

In order to approach knapsack bounds, score-based caching methods have been proposed, which select cache content based on a score $S_k$ for each object $O_k$. GreedyDual [1][2][17] [38][67][83][112] and score-gated caching strategies [14][15] [37][57][92][116] are aware of the relevant object properties for cache hit ratio and value optimization. More work in the same direction is denoted as utility maximization [34][93].

GreedyDual methods maintain a strictly sorted cache list due to scores [83] for storing data with highest value density in the cache, following the usual knapsack solution heuristics. As main drawback, cache updates involve sorting with high $O(\log M)$ effort, where new data is uploaded per cache miss.

Score-gated caching admits a requested external object only if its score exceeds the score of the eviction candidate(s) [58]. Score-gated clock (SGC) identifies the eviction candidate(s) by a clock scheme with fast $O(1)$ update speed, which even undercuts LRU for simple scores like $S_k = v_k/s_k$. The data upload churn of SGC is essentially lower than for GreedyDual and LRU, see Section 7.2. SGC selects and keeps top-scored content over time without strictly enforced sorting.

The utility optimization approach [93] achieves approximate score-based sorting similar to SGC. A simulated annealing algorithm with random eviction candidates and probabilistic insertion decisions is expected to get closer to an optimum static caching solution over time, on account of higher update effort and content variance for dynamic caching. The utility-based TTL approach [34] is mapping scores into TTL timers with similar effect as a direct or probabilistic preference.

In the evaluations of Figure 4, SGC is applied with score function $S_k = c_k/s_k$, with a request count $c_k$ replacing $p_k$ as score component of the bound. Hit ratio gaps of strategies without awareness of object properties are growing with the coefficient of variation of the underlying scores $S_k = v_k c_k/s_k$. Different object sizes $s_k$ and values $v_k$ add to variability as shown in the evaluations of Figure 4 and Figure 6 - Figure 9.

*3.5. Machine Learning Techniques*

Machine learning methods provide another class of score-based caching with even more flexibility to adapt to varying request pattern and object properties [107]. Neural network approaches are proposed for data on RAM, disk and virtual memory [71] as well as for web caching [28][44][74][104] [114][125]. Neural networks can optimize the hit ratio and other objectives by enhanced caching decisions. They implement score estimators via weight functions in multi-layer structures of neural nodes to predict the next requests [114].

Kirilin et al. [74] include the object size $s_k$, frequency count $c_k$, and 4 parameters on temporal and ordinal recency in their exponential smoothing factors as information base to optimize the hit ratio by machine learning. They propose a batch processing mode for reducing the computation effort in high request workloads. The update speed of neural network methods cannot be expected to cope with simple LRU or SGC caching principles. Moreover, long training phases are required to approach stable and performant machine learning results in a range of $4 \cdot 10^5 - 4 \cdot 10^7$ requests as indicated in [74].

Large performance gains of machine learning based caching are shown in [28][44][74][114][125] as compared to some state-of-the-art methods. However, simple score-based strategies are experienced to achieve fully competitive caching performance [28][44] without learning functions, when they share the same information about the request and object properties. The machine learning approaches are still at an early stage without fully exploiting all cache management options, with main focus on eviction/replacement policies [28][71][104][114], or on admission schemes [74]. Both, admission and eviction of cache content is considered in parallel but separated in [125], whereas score-based strategies like SGC directly compare the scores of admission and eviction candidates. An efficient ML adaptation approach is evaluated by [25], which adapts to LRU- and LFU-friendly workloads. Performance analysis of multi-layer machine learning algorithms is an open future research topic. Nonetheless, the knapsack bounds of Sections 3 and 4 are generally applicable for basic as well as advanced ML strategies.

## 4. GENERAL BOUNDS ON OPTIMUM CACHING

The previous bounds on hit and value ratios do not cover all types of correlated request sequences. When decisions on cache updates can anticipate the request sequence with knowledge about future requests, then Belady's algorithm [7][86][90] achieves the maximum cache hit ratio for arbitrary request pattern. Even if clairvoyant caching is unrealistic for most applications, Belady's upper bound is a common reference for caching performance. The bound indicates the potential for improvement by prediction or by using partial information about future requests [55]. In case of program code optimization, the data reference sequence is often known as a precondition to apply a clairvoyant caching strategy [68], but future requests are usually unknown in web caching.

Belady's algorithm follows a "*farthest next request first*" eviction principle. Therefore, objects in the cache are sorted due to the time until their next request. In case of a cache hit, the next request time and the implied new object rank is updated. Upon a cache miss, the requested object is inserted into the cache, if and only if its next request time comes before the next request time of a cached object. In this way, Belady's algorithm performs content selection based on the next request times of the objects as score criterion.

However, Belady's bound only applies for unit object size and for the hit count. Extensions for web caches with objects of different size and regarding caching values have not been addressed before 2018 [11][59], as to the authors' knowledge. Extended knapsack solutions to obtain maximum caching value are NP-hard, but standard knapsack heuristics are used to obtain upper and lower bounds around the maximum.

*4.1. 2D-Knapsack Solution for Maximum Caching Value*

We specify optimum clairvoyant caching with size and value per data as a 2-dimensional (2D-)knapsack solution [19][59]

- with a storage dimension limited by the cache size $M$ and
- with a time dimension, which refers to the index of the request sequence $r_1, r_2, ..., r_R$ of length $R$.

A request $r_n$ is a hit, if the requested object $O_{r_n}$ was stored in the cache during the inter-request interval $I = (m, n)$ since its previous request $r_m$ ($O_{r_m} = O_{r_n}$ and $O_{r_p} \neq O_{r_m}$ for $m < p < n$). Therefore, space must be blocked in the 2D-knapsack to realize a cache hit in the format of a rectangle, with size $n - m$ in time and $s_{r_n}$ in the storage dimension ($s_{r_n}$: size of $O_{r_n}$). In general, a rectangle $(n - m) \times s_{r_n}$ represents the required cache occupation to be placed in the 2D-knapsack per cache hit. Then the maximum hit count for clairvoyant caching is equal to the maximum number of rectangles which fit into the 2D-knapsack, as illustrated in Figure 5. For a more precise specification, let $I_{j,k}$ denote the $j^{th}$ request interval of an object $O_k$ in a request sequence, and let $\Lambda_{j,k}$ denote its length in time ($\forall j = 1, ..., R_k - 1; k = 1, ..., N$), where $R_k$ is the number of references to $O_k$ in a request trace. Then we have

$I_{j,k} = (m, m + \Lambda_{j,k} - 1); I_{j+1,k} = (m + \Lambda_{j,k}, m + \Lambda_{j,k} + \Lambda_{j+1,k} - 1);$
...; $O_{r_m} = O_{r_{m+\Lambda_{j,k}}} = O_k$; $O_{r_p} \neq O_k$ for $m < p < m + \Lambda_{j,k}$.

We also include the value $v_k$ of an object $O_k$. Let $C$ denote a set of intervals of valid object placements, which fit into the 2D-knapsack of total cache size $M \times R$ in space and time. Then the value of all placements is $V(C) = \Sigma_{j,k: I_{j,k} \in C} v_k$. Moreover, let $V_{max}$ denote the maximum achievable value, such that $\forall C: V(C) \leq V_{max}$.

Finally, the score function $S(I_{j,k}) = v_k/(\Lambda_{j,k} s_k)$ is introduced as value density of $I_{j,k}$, i.e., as the ratio of value to space occupation for a cache hit at the $(j+1)^{th}$ request to $O_k$. Then the usual knapsack heuristic is to sort all intervals $I_{j,k}$ due to their scores $S(I_{j,k})$ and to place as many as possible intervals into the 2D-knapsack starting from those with highest score. Note, that the placement of $I_{j,k}$ is fixed in time scale from $m$ to $m+\Lambda_{j,k} - 1$. One-timers or one hit wonders, i.e., objects with $R_k = 1$, can be ignored.

If an interval doesn't fit into the 2D-knapsack, we continue with intervals of next smaller value density $S(I_{j,k})$, until all of them are checked. In this way, a standard knapsack heuristic of valid content placements for a request sequence $r_1, r_2, ..., r_R$ and cache size $M$ is achieved, whose value $V^-$ is smaller than the maximum clairvoyant cache value ratio $V^- \leq V_{max}$.

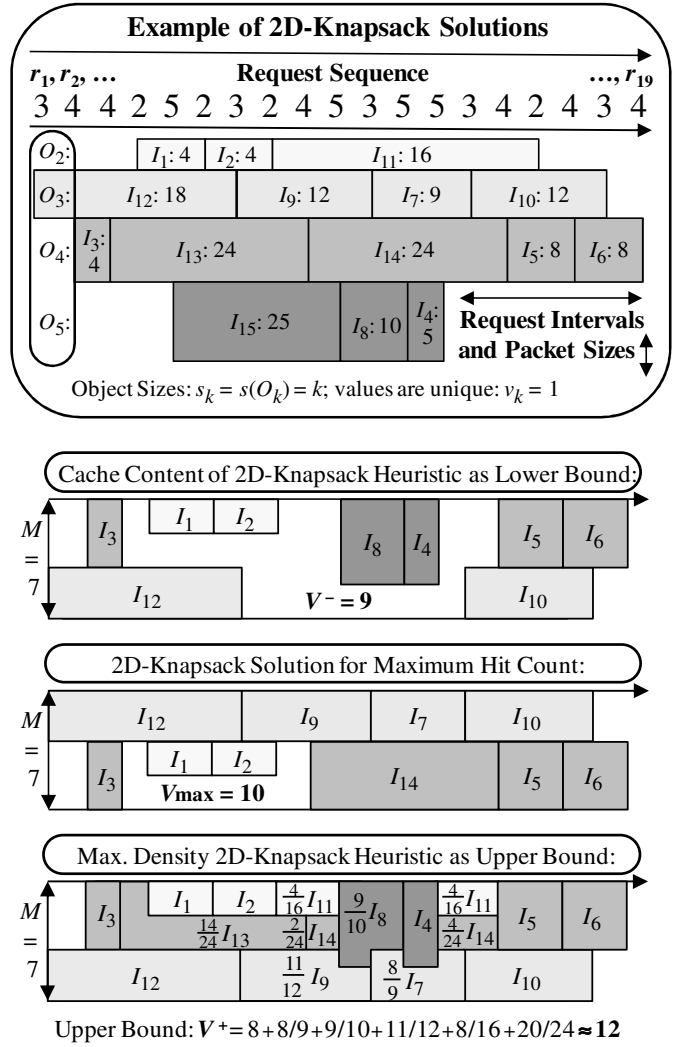

Figure 5: 2D-knapsack bounds for optimum caching

Figure 5 shows an example for computing the 2D-knapsack bounds around the maximum hit count. On top, a request sequence is shown with corresponding intervals $I_{j,k}$ and rectangles $\Lambda_{j,k} \times s_k$. The objects $O_2, ..., O_5$ have different sizes $s_k = k$ and unit value $v_k = 1$. We renumber the intervals $I_1 \cong I_{2,1}, I_2 \cong I_{2,2}, I_3 \cong I_{4,1}, ..., I_{15} \cong I_{5,1}$ in the order of increasing size $\sigma_m = \Lambda_{j,k} s_k$, such that $\sigma_1 = 2 \leq \cdots \leq \sigma_{15} = 25$ for $I_1, ..., I_{15}$, which is the order of decreasing value density.

In a next part below, the lower 2D-knapsack bound $V^-$ is shown. $V^- = 9$ intervals are found to fit into a cache of size $M = 7$, when placements are checked in the order $I_1, ..., I_{15}$. Next, a content placement for maximum hit count $V_{max} = 10$ is shown. Computation of the maximum is generally NP-hard.

*4.2. Upper 2D-Knapsack Bound of the Caching Value*

The knapsack solution for arbitrary object size and value can be modified for providing an upper bound $V^+$ of the optimum caching value $V_{max}$ as follows: Again we start with the previous 2D-knapsack heuristics for placing intervals in descending order of scores $S(I_{j,k})$. However, if an interval $I_{j,k}$ only partly fits into the 2D-knapsack, we don't ignore it, but we include a contribution by the part that fits. An area in the 2D-knapsack for the fitting part is blocked and a value $f * v_k$ is added to $V^+$, where $f^* \leq 1$ is the fraction of the size $\Lambda_{j,k} s_k$ of the interval $I_{j,k}$, which fits into the 2D-knapsack. In this way, an upper bound $V^+ \geq V_{max}$ of the maximum is obtained, as illustrated by the example at the bottom of Figure 5.

While the heuristic for placing only complete intervals $I_{j,k}$ into the knapsack leads to a lower bound $V^- \leq V_{max}$, the placements of all partially fitting intervals in the same order, fills every part of the 2D-knapsack with the maximum available value density $S(I_{j,k}) = v_k/(\Lambda_{j,k} s_k)$, thus yielding an upper bound $V^+ \geq V_{max}$. The bounds $V^-$ and $V^+$ provide a useful and tractable estimate $V^- \leq V_{max} \leq V^+$ of the optimum value $V_{max}$. The complexity for computing $V^-$ and $V^+$ is $O(R \cdot \max(M^*, \log R))$, where $R$ intervals are sorted due to scores $S(I_{j,k})$ and $M^*$ is the mean number of objects in the cache.

We need $M^*R$ updates of the cache occupation counters to put the intervals into the knapsack. A check, if an interval fits, has complexity $O(\log R)$. We update the knapsack with large sets of intervals in one step to speed up the computation.

### 4.3. Belady's Algorithm as a Special 2D-Knapsack Solution

In the special case of unit object size and value $s_k = v_k = 1$, the score function $S(I_{j,k}) = v_k/(\Lambda_{j,k} s_k) = 1/\Lambda_{j,k}$ depends only on the interval lengths. Then sorting due to $S(I_{j,k})$ means sorting due to increasing request interval lengths $\Lambda_{j,k}$, i.e., objects with shortest next request interval are preferred for caching. Obviously, this is equivalent to the opposite "*farthest next request first*" eviction principle of Belady's algorithm. Concluding, the knapsack solution includes Belady's algorithm as special case. In this case, the knapsack heuristic is exact and achieves the maximum hit count: $V^- = V_{max}$.

### 4.4. Application of the Bounds

We compute the knapsack bounds for optimum clairvoyant caching in examples with varying object sizes and values similar to the evaluations of Figure 3 - Figure 4. A trace of about 25 million requests to 1.5 million web objects is used, representing a one-day extract of an access trace from a large user population to web sites, which was also evaluated and described in more detail in [56]. 10% of the requests at the start and at the end are excluded from the evaluation to avoid initial cache filling phases, and because lower cache capacity is needed in final phases, when next request times are often beyond the end of the trace.

Figure 6 compares Belady's bound for unit size objects with FIFO, RANDOM, LRU, and score-gated clock as basic non-predictive strategies. Figure 7 - Figure 9 compare the cache hit ratios of this set of strategies with the 2D-knapsack bounds. The lower bound $V^-$ is marked by "×", and the upper bound $V^+$ by "+", respectively. They are computed by the algorithms for knapsack heuristics explained in Sections 4.1-4.2. SGC prefers objects according to the score $S_k = c_k v_k/s_k$, where $c_k$ is the current request count of an object $O_k$.

Another curve shows the 1D-knapsack solution of optimum static caching, based on a modified score function $S_k = (R_k - 1) v_k/s_k$. The factor $R_k - 1$ eliminates the influence of one-timers, where $R_k$ is the total request count of $O_k$ in the trace. However, static caching doesn't provide an upper bound for those non-IRM request traces, which exhibit moderate correlation between requests.

Figure 6 starts with objects of unit size and value $s_k = v_k = 1$. Then Belady's bound characterizes optimum clairvoyant caching and is equivalent to the 2D-knapsack heuristic. It exceeds the static caching results by about 5%. The cache hit ratio of score-gated clock and static caching achieve about the same hit ratio but with up to 15% gain over LRU. Similar hit ratio curves for LRU, FIFO and the bound are shown in a comparable example also in Figure 6 by Li et al. [81].

Figure 7 evaluates the value hit ratio for the same request trace with unit size objects, but their values $v_k$ are varied by a random generator with a lognormal distribution as defined in Eq. (6) with $\mu \approx -1.9$ and $\sigma = 1$. The parameter $\mu \approx -1.9$ is set for a normalized mean object value of 1. The cache hit ratio results of Figure 7 are similar to Figure 6, i.e. almost the same for LRU, FIFO and RANDOM methods, which are not aware of different object values. SGC with $S_k = c_k v_k/s_k$ prefers content of high value density and thus improves the value hit ratio up to 20% beyond LRU. Since Belady's bound can't include object value, we refer to the upper and lower 2D-knapsack bounds, with the optimum caching hit ratio $V_{max}$ nested between both bounds $V^- \leq V_{max} \leq V^+$. The clairvoyant capabilities achieve about 5% value hit ratio gain over SGC.

Figure 8 considers the case of unit object values, but different objects sizes $s_k$ according to the lognormal distribution defined in Eq. (6). The results for trace-based requests in Figure 8 are similar to those of Zipf distributed requests in Figure 4. The 2D-knapsack bounds of clairvoyant caching are again about 5% above the static caching results, which are close to the SGC results. Finally, Figure 9 shows results for objects of different sizes and different values for independent lognormal distributed object sizes $s_k$ according to Eq. (6) with $\mu = 3.5$ and $\sigma = 2.5$ and values $v_k$ again with parameters $\mu \approx -1.9$ and $\sigma = 1$.

The gaps between cache value ratios for LRU, FIFO, RANDOM and score-based caching are increasing with the variance in the scores $S_k = c_k v_k/s_k$ of the objects. As an extreme case included in Figure 9, the LRU strategy requires about 300 MB cache capacity in order achieve 10% cache value ratio, which is about 180-fold more than needed by SGC for the same value ratio. All three components of the SGC score function $S_k = c_k v_k/s_k$ contribute to the huge difference, where the mean size of objects stored by LRU is over 10-fold larger than for objects stored by SGC.

### 4.5. Min-Cost Flow Optimization and Tightness of the Bounds

In the example of Figure 5, the bounds are not very tight: $VHR^- = V^-/V_{Total} = 9/15$, $VHR^+ = V^+/V_{Total} = 12/15$ and $VHR_{max} = 10/15$. In Figure 7 - Figure 9, both 2D-knapsack bounds $VHR^+$ and $VHR^-$ differ by less than 2% for small caches and less than 2‰ for large caches in trace-driven evaluations for over a million objects. A statistical multiplexing effect leads to tighter bounds, with increasing cache size $M$ over a scale factor of $10^5$. In this case we can estimate the maximum hit ratio $VHR_{max} \approx (VHR^- + VHR^+)/2$ subject to 1% absolute deviation in relevant web cache scenarios.

Berger et al. [11] were the first to extend Belady's hit ratio bounds of clairvoyant caching for variable object sizes. They avoid knapsack approaches with a remark: "*... heuristics that work well on Knapsack perform badly in caching*", and derive min-cost flow optimization results for objects of different size and unit value. Similar to the knapsack bounds in Section 4.1, tractable upper and lower bounds are computed around the maximum hit ratio, whose computation is NP-hard.

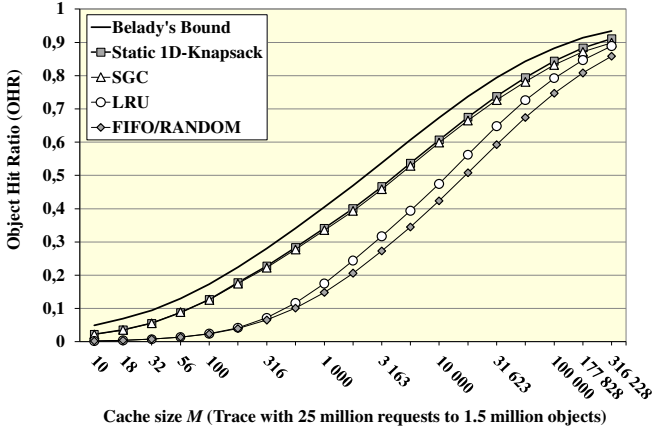

Figure 6: Cache hit ratios for unit object sizes and values

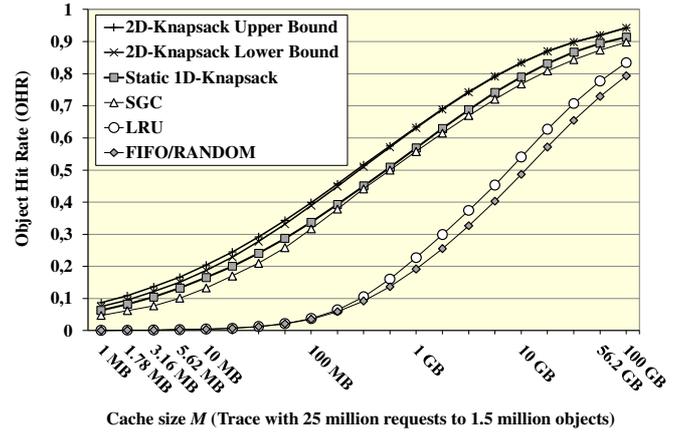

Figure 8: Cache hit ratios for objects of different size

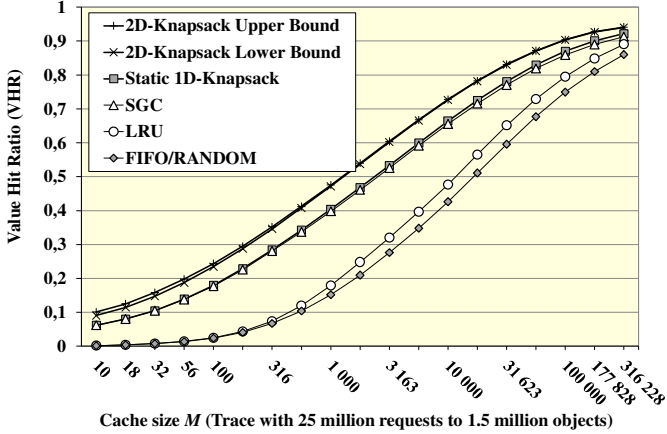

Figure 7: Cache value ratios for objects of different value

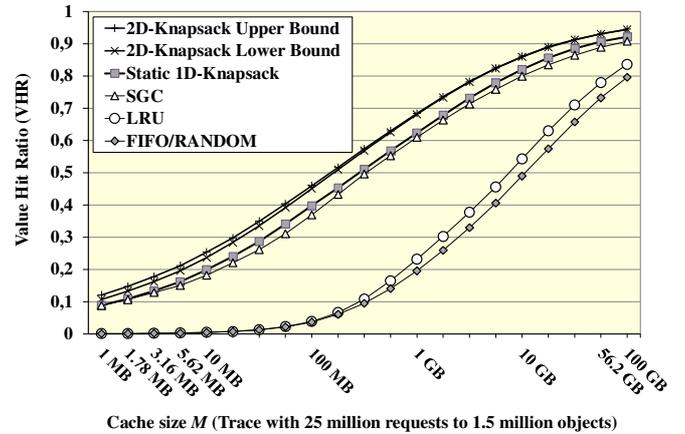

Figure 9: Cache value ratios for different sizes and value

One of both Practical Flow-based Offline Optimum (PFOO) bounds in [11], Section 6.1 follows the highest value density principle for the upper knapsack bound $VHR^+$ in a simpler, but less tight format: *"PFOO-L greedily claims the smallest intervals while not exceeding the average cache size."*, i.e. not exceeding $R \cdot M$.

The tightness of the min-cost flow optimization bounds [11] is in the same range of 1% difference as for knapsack bounds and partly even better. The approach [11] is restricted to variable object sizes $s_k$, whereas the standard knapsack approach also includes the values $v_k$ per object from the start for awareness of benefit, costs or other specific object properties. This makes score functions of GreedyDual and SGC strategies directly transferrable to corresponding knapsack weights, values and finally bounds.

*4.6. Data Retrieval with Delayed Hits and Limited Look-ahead*
Recent work by Atre et al. [3] is addressing the special aspect of how to handle delays after cache misses, when data is retrieved from a content server. The case of delayed hits is considered, when content is fetched from a server after a cache miss, such that new requests for the same content have to wait, while the retrieval is still ongoing. The delays for those waiting requests have to be considered by caching strategies and their evaluation, when delay reduction is a main caching goal. A caching schedule for minimum delay with regard to data retrieval scenarios has been derived by solving a min-cost multi-commodity flow problem [3][11].

On the other hand, retrieval delays can be exploited to postpone decisions about the eviction candidate to be replaced by the retrieved data. If an eviction candidate is requested during a retrieval delay, the decision can be changed to other candidates. Then a limited look-ahead into the request sequence can improve the hit ratio when eviction decisions are aware of the most recent requests. A moderate hit ratio improvement of a few percent is experienced due to such look-ahead in realistic web caching scenarios [55].

The study [3] shows that delayed hits during retrievals can have more significant effect on the overall delay savings of caches. A relaxed Belady machine learning algorithm [114] also classifies next requests before and beyond a look-ahead threshold denoted as Belady boundary.

An analysis study by Dehghan et al. [33] evaluates pending interest tables collecting requests experiencing retrieval delays for TTL caching in ICN networks. Cycles over 3 phases are considered. In the 1[st] phase, an object resides in the cache for a limited time-to-live, followed by a 2[nd] phase outside the cache, until a next request starts a new retrieval of the object in a 3[rd] phase, until the retrieval is finished. The work [33] is focused on the hit ratio rather than the delay. Only the 1[st] phase includes hits, whereas all requests waiting in the pending interest table are cache misses. Renewal and Poisson request processes are assumed per object, from which TTL approximations to LRU and FIFO caches are derived.

## 5. MARKOV ANALYSIS FOR INDEPENDENT REQUESTS

Markov analysis accompanies the evaluation of caching from the first proposals on LRU [72][73][88], FIFO, RANDOM [52] and optimum clairvoyant strategies [75][113], until extended results in recent time [8][12][50][54][78][105][110]. In this part, we outline the scope of exact steady state hit ratio solutions for basic caching strategies. The analysis of dynamically changing cache content is more complex than the static cache hit ratio formula of Eq. (1), and involves large Markov state spaces. Nonetheless, Markov analysis results are a versatile means for getting insights into caching performance. Extended solution formats are derived for multi-segment caches, for objects of different size and for networks of caches.

W.F. King [72] provided steady state IRM hit ratio formulas for LRU and FIFO. The FIFO result was confirmed also for RANDOM replacements by Gelenbe [52] and includes score-based strategies. In Section 5.1 we summarize the core proofs of those basic Markov caching results. Their verification via steady state equilibrium equations requires only a few lines, whereas the original proofs start from definitions for transient behavior and cover an A0 method in a broad framework over a journal and a long companion paper [52][73].

In the Sections 5.2 - 5.4, we show that the LRU steady state solution [72] also characterizes transient cache filling phases. This relationship is used to determine the LRU convergence or mixing time [78], which is equivalent to the notation of a *"characteristic time"* in approximation approaches of the LRU hit ratio [23][48]. An extension of the LRU results to objects of different size is provided in Section 5.5, whereas the common product form solution for FIFO, RANDOM and clock schemes as well as a general performance gain of LRU over FIFO [122] is shown to be restricted to unit size objects.

The common product form solution for FIFO, RANDOM and clock-based strategies has been extended to multi-segment caches by Gast and van Houdt [50][78]. Sections 5.6 - 5.8 briefly summarize this and further extensions to caches with probabilistic admission schemes and for generalized cache value and utility [34][92][93][115]. Markov analysis for clairvoyant caching is briefly addressed in Section 5.9 and time-based Markov approaches are presented in Section 6.

### 5.1. Product Form Solution of the Steady State IRM Hit Ratio for FIFO, RANDOM, and Basic Clock Strategies

*Stationary IRM Cache Content Solution for FIFO*

The FIFO cache eviction policy removes the object that has been in the cache for the longest time. It can be implemented by a stack or cyclic list. Upon a cache miss, the new requested item is put on top of the stack, replacing the object at the bottom. The FIFO cache remains unchanged for cache hits.

The Markov model for FIFO caches under IRM request pattern includes $N(N-1) \cdot \cdots \cdot (N-M+1)$ states for all combinations $O_{k_1}, \ldots, O_{k_M}$ of $M$ objects as cache content in the sequence of the FIFO stack. Then the steady state probabilities $p_{FIFO}(O_{k_1}, \cdots, O_{k_M})$ are obtained as a product form of the IRM request probabilities $p_{k_1}, \cdots, p_{k_M}$ [72] ($\forall j \neq l: k_j \neq k_l$)

$$p_{FIFO}(O_{k_1}, \cdots, O_{k_M}) = c\, p_{k_1} \cdot \cdots \cdot p_{k_M} \quad (8)$$

with normalization constant $c = 1 / \sum_{k_1, \ldots, k_M=1}^{N} p_{k_1} p_{k_2} \cdot \cdots \cdot p_{k_M}$.

For a proof of Eq. (8), the equilibrium equations of the Markov process are checked for each state transition per request. Since the FIFO cache content remains unchanged in case of cache hits, a state $(O_{k_1}, \ldots, O_{k_M})$ is preserved with probability $p_{k_1} + \cdots + p_{k_M}$. For a cache miss, transitions from the states $(O_{k_2}, \ldots, O_{k_{M-1}}, O_l)$ lead to $(O_{k_1}, \ldots, O_{k_M})$, where $O_{k_1}$ replaces $O_l$ with request probability $p_{k_1}$. Then the FIFO equilibrium equations for the steady state are obtained as ($\forall j \neq l: k_j \neq k_l$):

$$p_{FIFO}(O_{k_1}, \ldots, O_{k_M}) = (p_{k_1} + \cdots + p_{k_M})\, p_{FIFO}(O_{k_1}, \ldots, O_{k_M})$$
$$+ p_{k_1} \sum_{l=1;\, l \notin \{k_1, \ldots, k_M\}}^{N} p_{FIFO}(O_{k_2}, \ldots, O_{k_M}, O_l)$$

Substitution of the product form Eq. (8) confirms that this solution fulfills the equilibrium equations [52][73]:

$$c\, p_{k_1} \cdot \cdots \cdot p_{k_M} = (p_{k_1} + \cdots + p_{k_M})\, c\, p_{k_1} \cdot \cdots \cdot p_{k_M}$$
$$+ p_{k_1} \sum_{l=1;\, k_l \notin \{k_1, \ldots, k_M\}}^{N} c\, p_{k_2} \cdot \cdots \cdot p_{k_M} p_l \Leftrightarrow$$
$$1 = \sum_{l=1}^{M} p_{k_l} + p_{k_1} \sum_{l=1;\, l \notin \{k_1, \ldots, k_M\}}^{N} p_l / p_{k_1} = \sum_{l=1}^{M} p_{k_l} + \sum_{l=1}^{N} p_{k_l} - \sum_{l=1}^{M} p_{k_l} = 1.$$

In order to ensure convergence to steady state behavior of the confirmed product form solution, we finally have to check that the Markovian caching process is ergodic, such that a transition path leads from each state to each other with non-zero probability. States with less than $M$ objects are transient and only relevant during cache filling phases. The underlying IRM Markov chain for FIFO caches is generally ergodic except for the case $N = M + 1$. This result is included in criteria for ergodic caching networks derived by Rosensweig et al. [105].

The FIFO product form solution is derived in alternative ways via reversibility properties of the underlying Markov process by Cavallin et al. [22], Marin et al. [85] and via fluid flow approximation by Tsukada et al. [120].

*Stationary IRM Cache Content Solution for Clock per Request*

Clock caching methods [31] indicate an eviction candidate by a clock hand in a cyclic cache list. We consider a clock per request (CpR) scheme, which steps the clock at each request and, in case of a cache miss, replaces the object indicated by the clock hand with the requested one. Score-gated clock [58] combines a CpR scheme with score-based content admission. Corbato [31] already proposed CpR variants with improved hit ratio. Compared to CpR, the FIFO policy is similar and may be denoted as clock per cache miss. Steady state probabilities for CpR cache content $(O_{k_1}, \ldots, O_{k_M})$ with the clock hand pointing to the eviction candidate $O_{k_M}$ are obtained in the same product form as for FIFO:

$$p_{CpR}(O_{k_1}, \cdots, O_{k_M}) = p_{FIFO}(O_{k_1}, \cdots, O_{k_M}) = c\, p_{k_1} \cdot \cdots \cdot p_{k_M}. \quad (9)$$

The equilibrium equations of CpR differ from those for FIFO by a cyclic shift $(O_{k_1}, \ldots, O_{k_M}) \to (O_{k_M}, O_{k_1}, \ldots, O_{k_{M-1}})$ for cache hits, whereas the FIFO state stays unchanged after hits:

$$p_{CpR}(O_{k_1}, \ldots, O_{k_M}) = (p_{k_1} + \cdots + p_{k_M}) p_{CpR}(O_{k_2}, \ldots, O_{k_M}, O_{k_1})$$
$$+ p_{k_1} \sum_{\substack{l=1 \\ l \notin \{k_1, \ldots, k_M\}}}^{N} p_{CpR}(O_{k_2}, \ldots, O_{k_M}, O_l).$$

The product form solution of Eq. (9) again fulfills the CpR equilibrium equations, because the cyclic shift in the cache content does not change the state probability in product form.

*Stationary IRM Cache Content Solution for RANDOM*

A RANDOM strategy chooses an eviction candidate after cache misses with probability $1/M$ among the objects in the cache and leaves the cache unchanged after hits. Then the stack position of objects in the cache is not relevant and, in principle, a single Markov state is sufficient for all permutations of the same cache content. However, the product form of Eq. (8) assigns the same probability to permuted states and reveals to remain valid again for the RANDOM strategy [52] ($\forall j \neq l: k_j \neq k_l$)

$$p_{RANDOM}(O_{k_1}, \ldots, O_{k_M}) = c \, p_{k_1} \cdots p_{k_M}. \quad (10)$$

A state $(O_{k_1}, \ldots, O_{k_M})$ is entered for RANDOM evictions from all states, which differ in one object $O_l$ being replaced by $O_{k_j}$ with transition probability $p_{k_j}/M$. We obtain the equilibrium equations by summing up over all transitions to a state $(O_{k_1}, \ldots, O_{k_M})$ ($\forall j \neq l: k_j \neq k_l$):

$$p_{RANDOM}(O_{k_1}, \ldots, O_{k_M}) = (p_{k_1} + \cdots + p_{k_M}) p_{RANDOM}(O_{k_1}, \ldots, O_{k_M})$$
$$+ \sum_{j=1}^{M} \frac{p_{k_j}}{M} \sum_{\substack{l=1; l \notin \{k_1, \ldots, k_M\}}}^{N} p_{RANDOM}(O_{k_1}, \ldots, O_{k_{j-1}}, O_l, O_{k_{j+1}}, \ldots, O_{k_M})$$

The product form approach of Eq. (10) is again verified to fulfill the RANDOM equilibrium equations analogously to the FIFO and CpR cases. Therefore, the product form solution results in a common IRM hit ratio formula for FIFO, RANDOM and CpR, where the sum $p_{k_1} + \cdots + p_{k_M}$ of the request probabilities of all objects in the cache is included as the state specific hit ratio [72] ($\forall j \neq l: k_j \neq k_l$)

$$h_{IRM}^{FIFO} = h_{IRM}^{CpR} = h_{IRM}^{RANDOM} = \frac{\sum_{\substack{k_1, \ldots, k_M = 1 \\ k_1 < k_2 < \cdots < k_M}}^{N} p_{k_1} \cdots p_{k_M} \sum_{j=1}^{M} p_{k_j}}{\sum_{\substack{k_1, \ldots, k_M = 1 \\ k_1 < k_2 < \cdots < k_M}}^{N} p_{k_1} p_{k_2} \cdots p_{k_M}}. \quad (11)$$

The product form hit ratio result Eq. (11) can be evaluated for large $N$, $M$ in a recursive scheme with computational complexity $O(MN)$ as derived by Fagin and Price [43].

In general, product form solutions hold for a broader class of similar schemes. When we switch between FIFO, RANDOM and CpR in subsequent requests in a periodic or randomized sequence, then the equilibrium equations and ergodic behavior are preserved in many cases, whose full scope has to be settled in future work. Similar product form solutions are known for Markovian queueing networks, covering a wide range of different service classes [5], which are also adapted to cache nodes in the notation of Kelly cache networks [84].

*Comparing FIFO, CpR and RANDOM Caching Strategies*

Despite of a common steady state solution, the strategies can have different behavior for a specific request sequence. When, e.g., $M + 1$ different objects are periodically requested in a loop, the FIFO hit ratio is zero because the next request is always for the most recently evicted object, whereas the RANDOM hit ratio is $M/(M + 1)$. Such periodic request pattern may arise in caches for CPU paging workloads, whereas the differences between FIFO and RANDOM hit ratios are negligible in the previous evaluations of web request traces.

The IRM hit ratio of caching methods with product form solution is generally outperformed by LRU as proven in [122] and demonstrated in the evaluations of Figure 6 - Figure 9. Nonetheless, FIFO has applications as the caching scheme with highest update speed, see Section 7.1 [41].

### 5.2. IRM Hit Ratio for the LRU Caching Strategy

Equilibrium equations for LRU caches are more expansive [8], but the steady state cache content distribution under IRM conditions can be derived by simpler arguments. The top position of the LRU stack is occupied by the most recently addressed object $O_j$, i.e., $O_j$ is on top with probability $p_j$. The next request beforehand, which referred to another object, determines the object in the second LRU stack position. Then an object $O_k \neq O_j$ is found there with probability $p_k/(1 - p_j)$. In the third LRU stack position, another object $O_l \neq O_j$, $O_k$ is found with probability $p_l/(1 - p_j - p_k)$. We conclude that the steady state probabilities $p_{LRU}(O_{k_1}, \ldots, O_{k_M})$ for LRU cache content $(O_{k_1}, \ldots, O_{k_M})$ and the LRU hit ratio $h_{IRM}^{LRU}$ are given by [8][32][72][78][115] ($\forall j \neq l: k_j \neq k_l; \forall j: k_j \neq n$):

$$p_{LRU}(O_{k_1}, \ldots, O_{k_M}) = p_{k_1} \frac{p_{k_2}}{1 - p_{k_1}} \cdots \frac{p_{k_M}}{1 - p_{k_1} - \cdots - p_{k_{M-1}}}; \quad (12)$$

$$h_{IRM}^{LRU} = \sum_{k_1, \ldots, k_M = 1}^{N} p_{LRU}(O_{k_1}, \ldots, O_{k_M}) \sum_{j=1}^{M} p_{k_j} \quad (13)$$

$$= \sum_{n=1}^{N} p_n \left( p_n + \sum_{m=1}^{M-1} \sum_{k_1, \ldots, k_m = 1}^{N} p_{LRU}(O_{k_1}, \ldots, O_{k_m}, O_n) \right).$$

The latter representation of $h_{IRM}^{LRU}$ in the 2$^{nd}$ line of (13) distinguishes cases to find $O_n$ in the positions $m + 1 = 2, \ldots, M$ of the stack. The first term $p_n \cdot p_n$ represents the hit probability of $O_n$ in the top stack position. Note, that the probabilities $p_{LRU}(O_{k_1}, \ldots, O_{k_m})$ are valid not only for a set of $M$ objects that fills the cache, but also for stack rankings of the top $m$ objects on the entire range $m = 1, \ldots, N$ for any IRM request sequence with references to $m$ different objects. The evaluation assumes that only states with $M$ objects in the cache are relevant in steady state and involves $N!/(N − M)!$ summands.

## 5.3. LRU Cache Filling and Steady State Convergence Time

Li et al. [78] characterize the transient behavior of basic caching methods by mixing time estimates. They provide formulas for the order of magnitude of IRM mixing times of LRU, FIFO, RANDOM and CLIMB strategies in general and especially for Zipf distributed requests. LRU mixing times are shown to outperform FIFO, RANDOM and CLIMB. The mixing time results are extended to multi-segment caches [78].

Starting from an empty cache, LRU is already in steady state behavior as soon as the cache is filled.

The probability $p_{Cache\text{-}Fill}(O_{k_1}, \ldots, O_{k_m})$ that $O_{k_1}, \ldots, O_{k_m}$ are the first $m$ objects to enter an empty cache again follows the steady state result of Eq. (12-13) for the content distribution in an LRU stack. We obtain $p_{Cache\text{-}Fill}(O_{k_1}) = p_{k_1}$ for $O_{k_1}$ being the first object to enter an empty cache. When $O_{k_1}, \ldots, O_{k_{m-1}}$ are in the cache, $O_{k_m}$ will enter as the next object with probability $p_{k_m}/(1 - p_{k_1} - \cdots - p_{k_{m-1}})$. We conclude ($\forall j \neq l: k_j \neq k_l$)

$$p_{Cache\text{-}Fill}(O_{k_1}, \cdots, O_{k_m})$$
$$= p_{Cache\text{-}Fill}(O_{k_1}, \cdots, O_{k_{m-1}}) \cdot p_{k_m}/(1 - p_{k_1} - \cdots - p_{k_{m-1}}) \quad (14)$$
$$= p_{k_1} \cdot (p_{k_2}/(1 - p_{k_1})) \cdots \cdot p_{k_m}/(1 - p_{k_1} - \cdots - p_{k_{m-1}}) \Rightarrow$$
$$p_{Cache\text{-}Fill}(O_{k_1}, \cdots, O_{k_m}) = p_{LRU}(O_{k_1}, \cdots, O_{k_m})$$

i.e. when $m$ objects are in the cache during a filling phase, then the hit ratio of the next request equals the LRU hit ratio of Eq. (12-13) with cache size $m \leq M$. Thus, the LRU cache content distribution also characterizes cache filling phases. An alternative LRU cache content and miss ratio analysis in cache filling phases is proposed by [124] for IoT applications.

The cache filling behavior is still the same for other caching strategies such as FIFO, LFU, GreedyDual, score-based methods etc. [38][57][89][102], because they differ only in the treatment of evictions, but no evictions are performed during filling phases with enough room to add new objects.

In order to determine the LRU convergence time distribution, we define the probabilities $p_r(O_{k_1}, \ldots, O_{k_m})$ that the objects $O_{k_1}, \ldots, O_{k_m}$ have entered an initially empty cache during the first $r$ requests. We start from $p_1(O_{k_j}) = p_{k_j}$ for $j = 1, \ldots, N$. The next request leaves the set of cached objects unchanged in case of a cache hit or, after a miss, a new object enters. We compute $p_r(O_{k_1}, \ldots, O_{k_m})$ in an iterative scheme and finally obtain the distribution $\text{Prob}\{CT_{IRM}^{LRU} = j\}$ of the LRU convergence time $CT_{IRM}^{LRU}$, corresponding to a partial coupon collection process ($\forall j \neq l: k_j \neq k_l; m \leq M$) [60]:

$$p_{j+1}(O_{k_1}, \ldots, O_{k_m}) =$$
$$p_j(O_{k_1}, \ldots, O_{k_m}) \sum_{\ell=1}^{m} p_{k_\ell} + \sum_{\substack{k_m=1 \\ \forall j < m: k_m \neq k_j}}^{N} p_j(O_{k_1}, \ldots, O_{k_{m-1}}) p_{k_m}; \quad (15)$$

$$\text{Prob}\{CT_{IRM}^{LRU} = j\} = \sum_{\substack{k_1, k_2, \ldots, k_{M-1}=1 \\ \forall j \neq l: k_j \neq k_l}}^{N} \sum_{\substack{k_M=1 \\ \forall j < M: k_M \neq k_j}}^{N} p_{j-1}(O_{k_1}, \ldots, O_{k_{M-1}}) p_{k_M}.$$

The evaluation of (15) involves sets of up to $M$ cached objects and is tractable only for small caches.

For a simple and tractable estimate of the mean LRU convergence time $\overline{CT}_{IRM}^{LRU}$, we observe that an object $O_k$ is referenced within $r$ requests with probability $p_r(O_k) = 1 - (1 - p_k)^r$. Then a mean number $\#_{Objects}(r) = \sum_{k=1}^{N} 1 - (1 - p_k)^r$ of objects has entered an empty cache after $r$ requests. LRU is converging, when $\#_{Objects}(r)$ approaches the cache size $M$. We conclude

$$M \approx \#_{Objects}(\overline{CT}_{IRM}^{LRU}) = \sum_{k=1}^{N} 1 - (1 - p_k)^{\overline{CT}_{IRM}^{LRU}}. \quad (16)$$

The right side of Eq. (16) is monotonously increasing with $\overline{CT}_{IRM}^{LRU}$ from 0 to $N$, yielding a unique solution for $\overline{CT}_{IRM}^{LRU}$. The same format of Eq. (16) was proposed by Fagin [42] as the basis of an LRU hit ratio approximation, although not in the context of cache filling and convergence times. Che's approximation is similar, see Section 6.3.

The LRU steady state convergence time analysis holds for arbitrary starting conditions, because only the requests to the last $M$ different objects are relevant for the current LRU cache content status, independent from requests beforehand.

## 5.4. Comparison of LRU, FIFO and LFU Convergence Times

We compare the previous derivations of LRU cache filling processes with simulation results. Besides the LRU convergence time, our focus is on the development of the cache hit ratio in the filling phase and afterwards for LRU and other caching strategies. We evaluate an example via simulation for Zipf distributed IRM requests according to Eq. (2) with $\beta = 1$ for cache size $M = 1000$ and $N = 10^6$ objects. Figure 10 shows increasing hit ratios of the $r^{th}$ request in the filling phase. Each result represents the mean value of 1000 simulations. In the example, simulated cache filling phases and Eq. (16) yield the same mean length $\overline{CT}_{IRM}^{LRU} \approx 1501$ of filling phases. The other strategies show the same behavior as LRU in filling phases.

After the cache is filled, LRU is in steady state at a constant hit ratio level, whereas other methods pass through a second transient phase from LRU to their own steady state behavior.

While RANDOM and FIFO hit ratios decline from LRU level to a lower level, the LFU hit ratio is increasing towards the maximum IRM hit ratio in a long-lasting convergence phase, which is still about 1% below the maximum level after 50 000 requests [60]. The partition of the convergence time of FIFO, RANDOM and LFU policies in a first phase representing the LRU convergence time and an additional second transient phase is in line with mixing time estimates by Li et al. [78], which show that LRU adapts faster than FIFO, etc.

The maximum IRM hit ratio is shown by a horizontal dotted line. $h_{IRM}^{MAX} = p_1 + p_2 + \cdots + p_M$ is achieved, when $M$ most popular objects $O_1, \ldots, O_M$ are cached, assuming $p_1 \geq \ldots \geq p_N$.

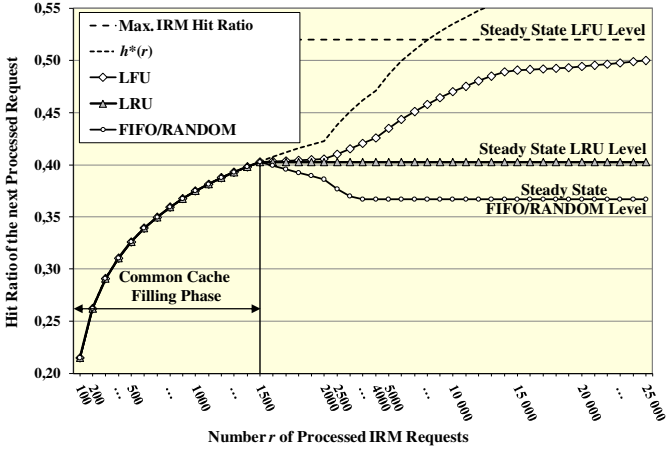

Figure 10: Hit ratio development in cache filling phases ($M = 1000$; $N = 10^6$; Zipf distributed requests with $\beta = 1$)

Moreover, Figure 10 shows a curve for a hit ratio bound $h^*(r)$ in the $r^{th}$ request. We compute $h^*(r) = \sum_k p_k (1-(1-p_k)^{r-1})$, where the term $1 - (1 - p_k)^r$ represents the probability that an object $O_k$ is requested and enters an empty LRU cache within the first $r$ requests. The bound $h^*(r)$ is exact, if no evictions are encountered, i.e. for $r \leq M$ or for caches of unrestricted size. As compared to the simulated LRU hit ratio curve, $h^*(r)$ has negligible deviations for $r < \overline{CT}_{IRM}^{LRU} \approx 1501$. For larger $r$, $h^*(r)$ increasingly overestimates the LRU hit ratio because of evictions that are ignored in the $h^*(r)$ computation [60].

### 5.5. Caching Performance with Objects of Different Size

For objects of different size, there is no 1:1 replacement. Upon a cache miss, eviction candidates are selected by the caching strategy (LRU, FIFO, RANDOM ...) until enough space is available for the new object. An object may be inserted without an eviction into free cache space, or several evictions may be required to insert a large object.

As to the authors' knowledge, extensions of the previous IRM hit ratio solutions for objects of varying size are not addressed in the literature. However, the LRU hit ratio result of Eq. (13) can be extended. Therefore, we consider an LRU cache for storing objects $O_k$ of different sizes $s_k$, where the fixed cache size $M$ is measured in Byte. Objects which do not fit into the cache are excluded, i.e., $s_k \leq M$ for $k = 1, ..., N$ and we still assume IRM requests with probabilities $p_k$. The requested object is always put on top of the LRU stack, while objects are evicted from the bottom if space is needed.

The steady state probabilities for the sequence of the top $m$ objects in an LRU stack are still valid for variable object sizes on the entire range $m = 1, ..., N$. Then the hit ratio in the format of the 2$^{nd}$ line of Eq. (13) can be straightforwardly extended, which distinguishes the contribution of each LRU stack position to the hit ratio by one summand, that can be restricted to sets of objects $O_k$ with size $s_k$ that fit into the cache. Then we obtain [60] ($\forall j \neq l: k_j \neq k_l; \forall j: k_j \neq n$):

$$h_{IRM}^{LRU} = \sum_{n=1}^{N} p_n \Big( p_n + \sum_{m=1}^{N-1} \sum_{\substack{k_1, ..., k_m = 1 \\ s_{k_1} + \cdots + s_{k_m} + s_n \leq M}}^{N} p_{LRU}(O_{k_1}, ..., O_{k_m}, O_n) \Big)$$

$$= \sum_{n=1}^{N} p_n^2 \Big( 1 + \sum_{m=1}^{N-1} \sum_{\substack{k_1, ..., k_m = 1 \\ s_{k_1} + \cdots + s_{k_m} + s_n \leq M}}^{N} \prod_{j=1}^{m} \frac{p_{k_j}}{1 - \sum_{i=1}^{j} p_{k_i}} \Big). \quad (17)$$

*No Common Product Form Solution for Varying Object Sizes*

Regarding the product form solution of Eq. (11) for FIFO, CpR and RANDOM strategies, we are not aware of hints in the literature about extensions for variables object size. We briefly show that the results are restricted to unit size objects.

We consider an example with $N = 3$ objects with request probabilities $p_1 = 0.2$; $p_2 = 0.3$; $p_3 = 0.5$, object sizes $s_1 = 1$; $s_2 = 2$; $s_3 = 3$ and cache size $M = 4$. Then two of the three objects fit together into the cache, except for $O_2$ and $O_3$ ($s_2 + s_3 = 5 > M$).

For this example, Markov chains with transitions corresponding to the LRU, FIFO, CpR and RANDOM strategies are shown in Figure 11. Evictions often leave a single object in the cache, when e.g. $O_2$ is requested while $O_3, O_1$ are in the cache. When $O_1$ is removed as first candidate, $O_3$ is also evicted to clear enough free space for $O_2$. Therefore, the steady state regime includes states for $O_3$ and $O_2$, whereas the state for $O_1$ is transient and will not be entered in steady state.

The steady state probabilities for each cache content are indicated in Figure 11, from which the IRM hit ratio is evaluated. In case of LRU, Eq. (17) provides a simpler alternative for direct computation of the hit ratio. Our check of simulation results for the example yields a perfect match, where confidence intervals indicate precision on 5 significant digits for $>10^9$ simulated requests. We obtain:

$$h_{IRM}^{LRU} = \frac{731}{1400} \approx 0.52214 < h_{IRM}^{FIFO} = \frac{131}{248} \approx 0.52823 < h_{IRM}^{CpR} = \frac{613}{1160} \approx 0.52845 < h_{IRM}^{RANDOM} = \frac{529}{1000} = 0.529. \quad (18)$$

$$BHR_{IRM}^{LRU} = \frac{3527}{6440} \approx 0.54764 < BHR_{IRM}^{FIFO} = \frac{3139}{5704} \approx 0.55032 < BHR_{IRM}^{CpR} = \frac{2937}{5336} \approx 0.55041 < BHR_{IRM}^{RANDOM} = \frac{2533}{4600} \approx 0.55065. \quad (19)$$

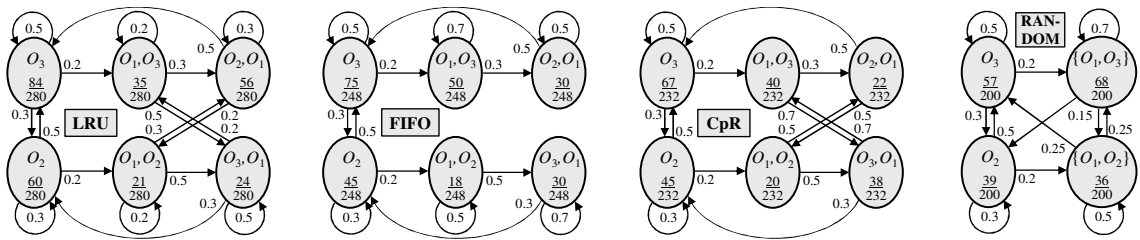

Figure 11: Markov chains and steady state probabilities in a small cache example with IRM requests and objects of different sizes

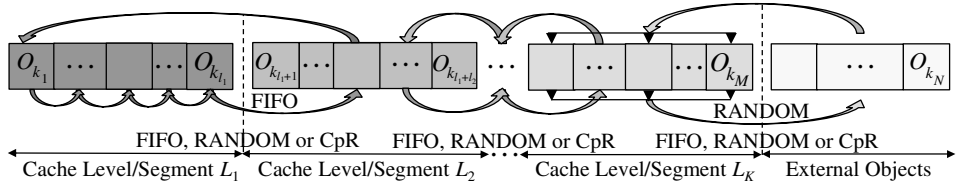

Figure 12: Multi-level/-segment caches with FIFO, RANDOM or CpR strategy for exchanging objects between the segments

The differences in the IRM hit ratio results in the example demonstrate that several well-known and proven properties for unit object size do not hold for objects of different size:

(1) The equivalence of FIFO, RANDOM and CpR hit ratios proven in [52] in a common product form solution of Eq. (11) are not valid for objects of different sizes.

(2) The "LRU is better than FIFO under IRM" result proven in [122] is also violated for objects of different sizes.

(3) Moreover, a monotonous increase of the hit ratio curves with the cache size $M$ is again valid only for unit object size, whereas zigzag shaped HRCs are encountered for LRU and FIFO with objects of different size [60].

On the other hand, simulation results of large caches for many objects of different size usually reveal negligible differences between FIFO, CpR and RANDOM, whereas the LRU performance can largely deviate from FIFO etc. in both directions. The preference for objects of higher popularity in LRU caches can lead to higher hit ratio, when those objects are small or, otherwise lower hit ratio when they are large. Score-based caching strategies [38][57] prefer objects $O_k$ with highest ratio $p_k/s_k$ for maximizing the caching performance, whereas the basic caching strategies are not aware of the object sizes and other relevant properties for optimization.

*5.6. Steady State IRM Hit Ratio for Multi-Level Caches*

On the other hand, the product form solutions of Section 5.3 can be extended to caches composed of several levels (lists, partitions, segments) $L_1, \ldots, L_K$. A requested object on cache level $L_j$ ($j = 2, \ldots, K$) is forwarded to level $L_{j-1}$ in exchange with an evicted object moving from $L_{j-1}$ to $L_j$ as illustrated in Figure 12. Many proposed caching strategies make use of two or more segments such as ARC [89], segmented LRU [102], $k$-LRU, LRU($m$) [51][78], FB-FIFO [53], half rank exchange [58], transposition relocation or CLIMB [88][115]. Distributed caching architectures also apply multi-level caching with commonly used storage being spread over several cache servers [6][78]. Multi-level caching is useful to improve the hit ratio by collecting most popular objects in the first cache levels on account of slower adaptation to content changes.

A steady state IRM hit ratio solution for $K$ level caches of different sizes $l_1, \ldots, l_K$ per level has been provided by Gast and van Houdt [50] for FIFO($K$) and RANDOM($K$) strategies. Li et al. [78] provides a solution for an LRU($K$) variant, which applies LRU on level $L_1$. Clock per request and combined variants of Section 5.1 can be included [60], even if different of those strategies are applied on different levels.

In general, the IRM steady state content distribution formula for $K$-level caches of different sizes $l_1, \ldots, l_K$ and arbitrary combinations of FIFO, RANDOM and CpR being applied on the levels has the extended product form [50] ($\forall j \neq l: k_j \neq k_l$):

$$p^*(k_1,\ldots,k_M) = c^* \prod_{j=1}^{K}(p_{k_{1+\sum_{i=1}^{j-1}l_i}} \cdot \cdots \cdot p_{k_{\sum_{i=1}^{j}l_i}})^{K+1-j}; \sum_{i=1}^{K} l_i = M; \quad (20)$$

$$h_{IRM}^{K\text{-Level Cache}} = \sum_{k_1,\ldots,k_M=1}^{N} p^*(k_1,\ldots,k_M) \sum_{j=1}^{M} p_{k_j} \Big/ \sum_{k_1,\ldots,k_M=1}^{N} p^*(k_1,\ldots,k_M).$$

This result includes half rank exchange [58] as a special case with segment sizes $l_j = 2^j$ and $M = 2^{K+1} - 1$. Climb [115] is covered by multi-level solutions [50][78] and in Eq. (20) as an extreme case with $\forall j: l_j = 1$. In one of the first Markov approaches for caching, McCabe [88] compared Climb, denoted as "*transposition relocation*", with LRU. The formula Eq. (20) is extensible to cases, when the cache size covers $k \leq K$ levels, such that $M = l_1 + \cdots + l_k$, and $K - k$ virtual levels are outside of the cache. The result Eq. (20) can be evaluated via a recursive scheme derived in [50] with computational complexity $O(N \cdot K^2 (l_1+1) \cdots (l_K+1))$.

The mixing time results [78] also include several cache levels. The adaptation and convergence time of cache content to changing request pattern can be significantly higher, when $K$ levels of equal size $M/K$ are used, such that only a fraction $1/K$ of the cache can be entered by a requested external object. The half rank exchange method [58] opens half of the cache size directly for external objects upon their next request with smaller segments for collecting popular objects in the other half. Studies on segmented LRU experience only minor improvements for more than two segments [53][89].

*5.7. Extended Markov Solutions for Probabilistic Caching*

Starobinsky and Tse [115] further extend the previous Markov solutions to probabilistic admission schemes. Therefore, each request for an object $O_k$ is either treated with probability $q_k$ by a predefined caching strategy, e.g. LRU, or otherwise the cache remains unchanged. Then $q_k$ can be used to prefer objects according to a score function, where $q_k = v_k/s_k$ is explicitly proposed in [115] according to the score applied in Eq. (5) for an optimized knapsack solution.

Starobinsky and Tse [115] show that probabilistic caching leads to an extended Markov solution in case of the LRU and the Climb strategy [88] by substituting request probabilities $p_k$ with factors $\gamma_k = p_k q_k / \Sigma_k p_k q_k$. Then Eq. (12) is extended to

$$p_{LRU-PC}(O_{k_1},\cdots,O_{k_M}) = \gamma_{k_1} \frac{\gamma_{k_2}}{1-\gamma_{k_1}} \cdots \frac{\gamma_{k_M}}{1-\gamma_{k_1}-\cdots-\gamma_{k_{M-1}}}$$

for probabilistic LRU caching. The conclusions of [115] also mention an extended FIFO solution. In fact, all product form solutions for FIFO, CpR, RANDOM and combined strategies of Section 5.3 can be generalized for probabilistic caching

$$p_{FIFO\text{-}PC}(O_{k_1}, \cdots, O_{k_M}) = p_{CpR\text{-}PC}(O_{k_1}, \cdots, O_{k_M}) = c\, \gamma_{k_1} \cdot \cdots \cdot \gamma_{k_M}.$$

Moreover, the generalization still fully applies to the product form solution for multi-segment caches of Eq. (20), as suggested by the proof for Climb as a special case in [115].

Another probabilistic $q_k$-LRU approach is proposed by Neglia et al. [92], which puts a requested external object into the cache with probability $q_k = e^{-\beta s_k/v_k}$ ($\beta > 0$) and uses LRU for evictions. In this way, objects are again preferred according to a score $S_k = v_k c_k / s_k$. Strict preference is enforced for $\beta \to \infty$ on account of reduced convergence and adaptation speed due to small update probabilities $q_k$. A modified format $q_k = n^{-\beta s_k/v_k}$ is recommended as enhanced Dynq-LRU method in [93], yielding favourable performance results in comparison to GreedyDual methods. $q_k$-LRU and Dynq-LRU are further enhanced by simulated annealing [92][93] to approach the knapsack bound of Eq. (5) by storing the highest scored content as also achieved by score-gated methods.

Although probabilistic caching can prefer most relevant cache content with flexible soft enforcement, a strict preference of score-based content selection methods [17][57][67] often leads to higher steady state hit ratio close to cache performance bounds. Moreover, probabilistic caching has longer adaptation and convergence times, since cache updates are omitted with probability $1 - q_k$, even if a scaling factor ensures that $\max_k q_k = 1$ [115]. In addition to probabilistic decisions for admitting objects to the cache, decisions about evictions can also be made probabilistic, where the effect on transient and steady state solutions is for future study.

*5.8. Extension to Value and Byte Hit Ratios*

Steady state Markov results are usually derived for the hit ratio. The cache value ratio is obtained as ($\forall j \neq l : k_j \neq k_l$)

$$VHR_{IRM} = \sum_{k_1,\ldots,k_M=1}^{N} p(O_{k_1}, \ldots, O_{k_M}) \sum_{j=1}^{M} p_{k_j} v_{k_j} \bigg/ \sum_{i=1}^{N} p_i v_i,$$

in extension of Eq. (4), where $p(O_{k_1}, \cdots, O_{k_m})$ is the steady state probability of cache content for a considered strategy. The caching value $v_k$ of an object $O_k$ can follow goals for optimized delays or network load etc., as discussed in Section 2. The result includes the Byte Hit Ratio, when the object size in Byte is taken as the value ($v_k = s_k$). In the example of Figure 11, we obtain the result in Eq. (19).

*5.9. Markov Analysis for Clairvoyant Caching*

Markov analysis applies to clairvoyant caching as an alternative to Belady's algorithm with about the same computation effort [54]. The Markov process for arbitrary request traces has $(M+1)^{(N-M)} N! / (N-M)!$ states in reduced state space compared to even larger initial models [75][113]. For the IRM case, a simple hit ratio formula is derived for $M = 1$ [54], but the evaluations soon become intractable for larger $M, N$.

## 6. TIME TO LIVE CACHING

*6.1. Basic Assumptions and Applications for TTL Caching*

Time-to-live caching policies assign a time limit to each object for control of valid cache content. When an object enters the cache, it is associated with a timer and will be invalidated when its timer expires. The expiry deadline can be set until the current content is expected to become stale. In this way, TTL constraints enforce a weak form of data consistency, such that invalid data, which has been updated or removed from a web content server will soon be also removed from the caches [18][29]. In fixed size caches, additional TTL restrictions can be combined with any usual cache management method (LRU, LFU, ML, etc.) [46].

On the other hand, pure TTL caches are assumed to provide enough storage for all valid content. Then TTL values per data can be adapted for control of the amount of valid cache content instead of data consistency. When TTL caches exclude expired data then shorter TTL values reduce the sojourn time of data in the cache and the entire amount of valid data in a TTL cache, as well as the hit ratio. Upon a next request, data can reenter the cache with a renewed TTL value.

Such adaptive TTL caching approaches were already proposed by Colajanni and Yu [30] for load balancing among DNS servers, and were investigated for many other purposes [6][12][20][23][26][27][34][45][49][51][66][76][91][108][126]. TTL caches are most useful for small and often changing web objects. They are deployed

- for the Domain Name Service [26][30][69][91],
- for search engine support in web query result caches [108],
- for caches in data centers of huge storage size as part of CDN [6] and cloud infrastructures [20][109] etc.

A trace-based study of DNS caches by Jung et al. [69] assumes a unique TTL timer for all records, which is varied in the range from seconds to one day. Their evaluation of a 24-hour TTL timer leads to 97% hit ratio, which is decreasing down to 80% for a 15 minutes TTL timer. TTL timers of the length of hours or one day are also recommended by Moura et al. [91] to reduce latency in DNS replies, if security and load balancing demands do not imply shorter TTL expiry.

DNS services and web searches are time critical applications with demands for fast responses, but with only small size data of a few bytes per record. Then sufficient cache space can be provided for covering all relevant data records. TTL approaches for caches in large-scale data centers often assume unlimited extensible storage that can be adapted to variable workloads with costs proportional to the storage space [20]. However, in practice, several storage types ranging from very fast, expensive and therefore small storage units to large and cheap mass storage are involved to optimize caching infrastructures in order to cope with high request workloads.

We notice that the behaviour and analysis of pure TTL caches is fundamentally different from strategies for selecting and replacing content in caches of fixed size, such that

- the request time instants are relevant for evaluating TTL cache hits and misses, whereas cache performance without TTL conditions can be evaluated based on the entire request sequence of all cacheable objects, but without regarding inter-request and other timing information,
- the presence of an object in a TTL cache depends only on its own TTL timing and is independent of other objects.

The first property requires TTL analysis to involve request timing per object, via Poisson, renewal, (semi-)Markov or general point processes for characterizing the request instances and inter-request times.

The independence property essentially simplifies the TTL cache performance analysis by enabling separated treatment per data object independent of the others [12]. This avoids huge Markov state spaces for representing the complete cache content. Such TTL analysis approaches are transferred to LRU, FIFO and some other policies for fixed size caches via Che's and Fagin's approximations [23][43] with approved high precision, as addressed in Section 6.3.

LFU, GreedyDual, and ML caching decisions introduce score-based comparisons and/or sorting of data objects. Then the precondition of an independent treatment per data object is violated as the basis for a simplified TTL analysis. An approach for reducing the variance of the amount of valid data in a pure TTL cache [13] leads to closer adaptation to a fixed cache size limit, but involves segmented caching.

In Section 6.2, basic TTL cache performance modeling and analysis is summarized and transferred to LRU and FIFO approximations in Section 6.3. Extended TTL results for correlated request pattern and caching networks are addressed in Sections 6.4 - 6.5. Finally, TTL constraints for data consistency are integrated into policies for caches of fixed size and bounds for optimum caching in Sections 6.6 - 6.7.

*6.2. TTL Cache Hit Ratio Analysis for Poisson Requests*

In general, the impact of TTL timers on the hit ratio can be analyzed starting from the simple assumption of Poisson processes for the requests per object, which imply IRM references for the total request stream to a cache.

*Poisson Process Model for the Requests per Object*

The Poisson reference model (PRM) for a fixed set of objects $O_k$ ($k = 1, …, N$) assumes the request count $c_k(\Delta T)$ of $O_k$ in a time interval $\Delta T$ to be Poisson distributed with rate $\lambda_k$:

$$\text{Prob}\{c_k(\Delta T) = m\} = (\lambda_k \Delta T)^m\, e^{-\lambda_k \Delta T}/m!$$

As a first consequence of the PRM assumption, an inter-request time $T_k$ between two consecutive requests to the object $O_k$ is exponentially distributed:

$$\text{Prob}\{c_k(\Delta T) = 0\} = e^{-\lambda_k \Delta T} \Leftrightarrow \text{Prob}\{T_k \leq \Delta T\} = 1 - e^{-\lambda_k \Delta T}.$$

The PRM properties are valid also for the total request stream of all objects, whose rate $\lambda$ is the sum of the request rates $\lambda_k$ per object. Poisson processes are memoryless, such that the entire request sequence follows IRM conditions with fixed probabilities $p_k$ that the next request refers to an object $O_k$:

$$p_k = \lambda_k/\lambda \text{ for } k = 1, …, N; \quad \lambda = \lambda_1 + \cdots + \lambda_N.$$

*TTL Control Options and Corresponding Hit Ratios*

Two cases can be distinguished for TTLs under cache control, as illustrated in Figure 13 [6][24][26][27][34][49][101]:

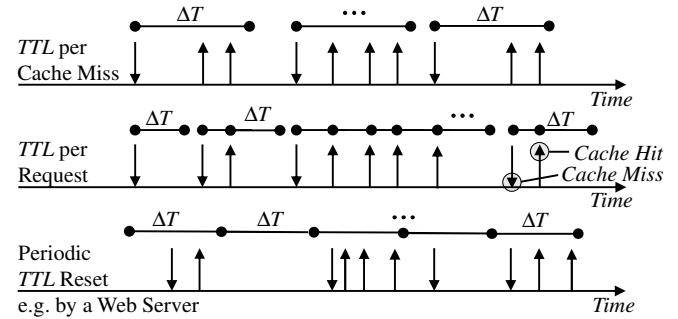

Figure 13: TTL reset options for caches and web servers

- *TTL Reset per Cache Miss*

The TTL is reset per cache miss, i.e., the object data and its TTL is renewed at the next request after the TTL has expired. The PRM hit ratio $h_{PRM}^{TTL,Miss}$ is then increasing with the mean number of hits $E[c_k(\Delta T)]$ on $O_k$ during the TTL interval $\Delta T$:

$$h_{PRM,O_k}^{TTL,Miss}(\Delta T) = \frac{E[c_k(\Delta T)]}{E[c_k(\Delta T)] + 1} = \frac{\lambda_k \Delta T}{\lambda_k \Delta T + 1}. \quad (21)$$

- *TTL Reset per Request*

When the TTL $\Delta T$ is reset per request then the next request to the object $O_k$ is a hit, if it arrives before TTL expiry:

$$h_{PRM,O_k}^{TTL,Req}(\Delta T) = \text{Prob}\{T_k \leq \Delta T\} = 1 - e^{-\lambda_k \Delta T}. \quad (22)$$

In both cases of Eq. (21 - 22), the hit ratio is monotonously increasing with $\Delta T$, as well as the expected number of valid objects and the required cache size to store them. Therefore, the TTL control can be adapted to a predefined target of the hit ratio or for the cache size [34].

- *Periodic TTL Resets*

With regard to data consistency, we also consider TTL under control of web servers as a third basic option. Then TTLs are triggered by updates of original data on the server independent of request instances. The TTLs can be forwarded together with updates of objects or in periodical exchange of control data between a web server and cooperating caches [18][46][129]. When the TTL has expired, the next request is a cache miss. Then the object is reloaded with the current TTL timer of the web server, i.e., without TTL reset. Upon a hit, data is delivered from the cache again with unchanged TTL timer. The diagram on the bottom of Figure 13 illustrates cache hits and misses for periodic TTL resets. An analysis of the hit ratio for periodic TTL intervals $\Delta T$ is based on only two parameters of the request process:

(1) the mean number $E[c_k(\Delta T)]$ of requests to $O_k$ in $\Delta T$, and
(2) the fraction $p_k^0(\Delta T)$ of intervals without a request to $O_k$.

There is one cache miss per interval with at least one request. Consequently, the mean number of TTL cache hits per interval is $E[c_k(\Delta T)] - 1 + p_k^0(\Delta T)$. We obtain the object specific cache hit ratio $h_{O_k}^{TTL,PR}(\Delta T)$ for periodic TTL resets:

$$h_{O_k}^{TTL,PR}(\Delta T) = (E[c_k(\Delta T)] - 1 + p_k^0(\Delta T))/E[c_k(\Delta T)]. \quad (23)$$

This result holds independent of request instances and thus applies generally without the need to assume special request timing models such as PRM as the basis of Eq. (21 - 22). On the whole, the cache hit ratio is obtained as the mean number of hits divided by the mean number of requests over all objects per TTL interval, again for arbitrary request pattern:

$$h^{TTL,PR}(\Delta T) = \sum_k (E[c_k(\Delta T)] - 1 + p_k^0(\Delta T))/\sum_k E[c_k(\Delta T)]. \quad (24)$$

The result further extends to individual TTL timers $\Delta T_k$ for each object $O_k$, when we consider mean values in the long term limit for the hit rate and the total request rate

$$h^{TTL,PR}(\Delta T_k) = \frac{\sum_k (E[c_k(\Delta T_k)] - 1 + p_k^0(\Delta T_k))/\Delta T_k}{\sum_k E[c_k(\Delta T_k)]/\Delta T_k}. \quad (25)$$

For a considered request trace, it is straightforward to evaluate both values $E[c_k(\Delta T)]$ and $p_k^0(\Delta T)$ for different TTL intervals $\Delta T_k$ and to adapt the TTLs for a target hit ratio or cache size similar to the approaches in [6][34]. Finally, the hit ratio is computed. In case of PRM we obtain

$$E[c_k(\Delta T)] = \lambda_k \Delta T_k; \quad p_k^0(\Delta T) = e^{-\lambda_k \Delta T_k} \Rightarrow$$
$$h_{PRM}^{TTL,PR}(\Delta T_k) = \sum_k [\lambda_k + (e^{-\lambda_k \Delta T_k} - 1)/\Delta T_k]/\sum_k \lambda_k. \quad (26)$$

*6.3. Approximations of the LRU and FIFO Hit Ratio*

An LRU hit ratio approximation for caches of fixed size $M$ has been derived via TTL analysis by Che et al. [23][48]. It is based on a unique TTL timer for all objects. The TTL approach to an LRU cache for $M$ objects of equal size adapts the timer $\Delta T_{LRU}$ to match the characteristic time [23] from a request instant that puts an object to the top LRU position until eviction. Starting from the top, the time $\Delta T_{LRU}$ until eviction equals the time until $M$ requests to different other objects are encountered. The PRM probability that $O_j$ is requested during $\Delta T_{LRU}$ is given by $1 - e^{-\lambda_j \Delta T_{LRU}}$, similar to Eq. (22). Then the mean number of requested objects during $\Delta T_{LRU}$ is the sum of those probabilities, which should be equal to $M$ [23][48]:

$$E[\#_{\text{Requested Objects}}(\Delta T_{LRU})] = \sum_{j=1}^N 1 - e^{-\lambda_j \Delta T_{LRU}} \approx M. \quad (27)$$

In principle, the computation of $\Delta T_{LRU}$ for a specific object $O_k$ should exclude the term for $j = k$ from the sum. However, the estimator turns out to be fairly accurate with all terms included, leading to lower computation effort for a common characteristic time $\Delta T_{LRU}$ for all objects.

In the next step, the hit ratio per object $O_j$ can be computed via the PRM result for TTL resets per request of Eq. (22):

$$h_{LRU}(O_j) \approx \text{Prob}\{T_j \leq \Delta T_{LRU}\} = 1 - e^{-\lambda_j \Delta T_{LRU}}.$$

In summary, Che's approximation of the hit ratio $h_{LRU}$ is computed in two steps [23][48] ($\Delta_{LRU} = \lambda \Delta T_{LRU}$ and $p_j = \lambda_j/\lambda$):

- First, the solution $\Delta_{LRU}$ of the equation

$$M = \sum_{j=1}^N 1 - e^{-p_j \Delta_{LRU}} \quad (28)$$

is determined. The sum is monotonously increasing with $\Delta_{LRU}$ from 0 to $N \geq M$, yielding a unique solution $\Delta_{LRU}$.

- Then the LRU hit ratio is obtained per object and in total:

$$h_{Che}(O_j) = 1 - e^{-p_j \Delta_{LRU}}; \quad h_{Che} = \sum_{j=1}^N p_j(1 - e^{-p_j \Delta_{LRU}}). \quad (29)$$

Finally, we note that the characteristic time $\Delta T_{LRU}$ as introduced by Che et al. [23] is equivalent to the LRU convergence time $CT_{IRM}^{LRU}$ as evaluated in Eqs. (15 - 16). Fagin [42] already proposed an approximation $h_{CT}$ of the LRU hit ratio based on Eq. (16) in a notation of "*expected working set size*":

$$M = \sum_{k=1}^N 1 - (1 - p_k)^{\overline{CT}_{IRM}^{LRU}}; \quad h_{CT}(O_k) = 1 - (1 - p_k)^{\overline{CT}_{IRM}^{LRU}};$$
$$h_{CT} = \sum_{k=1}^N p_k h_{CT}(O_k) = \sum_{k=1}^N p_k(1 - (1 - p_k)^{\overline{CT}_{IRM}^{LRU}}). \quad (30)$$

The convergence time approach of Eq. (16) differs from Che's approach by the factor $1 - p_k$ being substituted for $e^{-p_k}$ in Eqs. (28 - 29). The approach of Eq. (30) is exact for $M = 1$. We have compared the accuracy of both approaches in a detailed quantitative study [60], which strongly suggests that

- the maximum deviations $|h_{Che} - h_{LRU}|$ of Che's approximation are decreasing with the cache size $M$ from 8.25% for $M = 1$ downto less than 1% for $M \geq 10$;
- the maximum deviations $|h_{CT} - h_{LRU}|$ of Fagin's approximation are decreasing with the cache size $M$ from 5.2% for $M = 2$ downto less than 1.3% for $M \geq 10$.

However, a proof of those properties, which confirm generally good accuracy of both approximations in usual caching scenarios, is for future study. Moreover, both approximations are proven to become asymptotically exact for increasing cache size [9][48][65][66][96][103], but without providing concrete bounds on the accuracy. The study [60] also proposes an extension of the approximations for objects of different size. Then additional deviations are caused by a fraction of unused cache space, which can be estimated and reduced. Therefore, deviations can again be large for small caches, whereas asymptotically exact behavior is expected for large caches due to a statistical multiplexing effect, provided that all object sizes are becoming tiny compared to the cache size.

The TTL approximation of Eq. (28 - 29) is extended by Mazziane et al. [87] to obtain the hit ratio of an LRU variant for similarity caching, where data is updated also in case of requests for similar data within a limited distance according to a similarity measure. Similarity caching is useful when requests follow recommendation systems.

*Approximation of FIFO, CpR and RANDOM Cache Hit Ratios*

The TTL-based LRU hit ratio approximation of Eqs. (28-29) under a PRM and/or IRM assumption can be transferred to FIFO. Therefore we follow the derivation by Dehghan et al. [34] and Garetto et al. [49] as well as an equivalent previous approach by Dan and Towsley [32].

An object $O_j$ encounters a cache miss at each time instant when it (re-)enters the cache. Afterwards, a mean number $\lambda_j \Delta T_{\text{FIFO}}$ of hits is following during the next sojourn time of $O_j$ in the cache before eviction. We again assume a common FIFO cache sojourn time $\Delta T_{\text{FIFO}}$ for all objects.

Then the PRM hit ratio $h_{\text{FIFO}}(O_j)$ per object is obtained from Eq. (21): $h_{\text{FIFO}}(O_j) = \lambda_j \Delta T_{\text{FIFO}}/(\lambda_j \Delta T_{\text{FIFO}} + 1)$. For a transfer of the continuous time PRM assumption to a discrete time system, we reuse the probabilities $p_k = \lambda_k/\lambda$ for $k = 1, \ldots, N$ ($\lambda = \lambda_1 + \cdots + \lambda_N$) and we refer to $\Delta_{\text{FIFO}} = \lambda \Delta T_{\text{FIFO}}$ as the mean number of requests in a FIFO cache sojourn time $\Delta T_{\text{FIFO}}$. This finally leads to an IRM hit ratio approximation for FIFO corresponding to Che's LRU approach of Eqs. (28-29):

- First, the solution $\Delta_{\text{FIFO}}$ of the equation
$$M = \sum_{j=1}^{N} h_{\text{FIFO}}(O_j); \quad h_{\text{FIFO}}(O_j) = p_j \Delta_{\text{FIFO}}/(p_j \Delta_{\text{FIFO}} + 1) \quad (31)$$
is determined. The sum is monotonously increasing with $\Delta_{\text{FIFO}}$, yielding a unique solution $\Delta_{\text{FIFO}}$.

- Finally, the FIFO hit ratio approximation is given by
$$h_{\text{FIFO,Approx.}} = \sum_{j=1}^{N} p_j h_{\text{FIFO}}(O_j). \quad (32)$$

The accuracy of the FIFO approximation behaves similar to the LRU case, as shown in a quantitative evaluation [60] confirming less than 3% deviation for $M \geq 10$. The result also holds for RANDOM and CpR caching strategies because of their common steady state IRM hit ratio. However, the exact FIFO, CpR, RANDOM product form solution is scalable for large caches in contrast to the exact LRU solution. Thus, the FIFO approximation is less relevant for unit object size, but a useful basis for the extension to objects of different size [60].

*TTL Analysis for Generalized Utility and Score Functions*
The previous TTL-based approximation schemes cover a set of basic caching methods. Dehghan et al. [34] propose to adapt TTLs to maximize utility functions. In general, utility functions can be adapted with regard to fairness, throughput, delay, costs etc. per data object, similar to score-based caching, as addressed in Section 3.4. Utility functions corresponding to LRU, FIFO and to fairness principles are derived in [34], while a utility-based alternative to GreedyDual and score-based caching is implemented with a simulated annealing algorithm by Neglia et al. [93]. Another work by Dehghan et al. [33] extends the TTL analysis with regard to retrieval delays, as discussed in more detail in Section 4.6.

*6.4. TTL Modelling and Analysis for Correlated Requests*
The analysis of TTL caching performance has been extended to renewal processes with arbitrary inter-request time distribution Prob($T_m \leq \Delta T$) [49][69], to Markov renewal request processes [6][12][26][27][45][51], and to general ergodic point processes as independent request streams per object $O_m$ with utilities per content [98]. Those models imply correlated requests beyond PRM and IRM. Figure 14 illustrates extended modeling schemes. The second case of renewal inter-request times already can specify alternating phases of high and low request rates. Garetto et al. [49] extend the TTL-based LRU and FIFO approximations for renewal requests per object and for variants with probabilistic insertion, multi-segment caches, as well as for caching network scenarios.

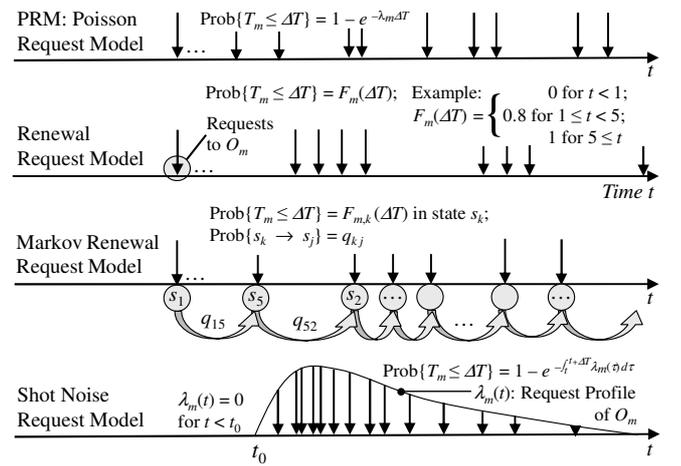

Figure 14: Poisson, Markov Renewal and Shot Noise Models

Markov renewal processes associate a state $s_k$ with each request, such that the next inter-request time for $O_m$ can follow different distributions $F_{m,k}(t)$ depending on the current state $s_k$. For PRM as well as for Markov renewal models, the analysis of Eq. (21 - 26) is still valid for all cases of $h^{TTL,Miss}(\Delta T)$, $h^{TTL,Req}(\Delta T)$, and $h^{TTL,PR}(\Delta T)$. The evaluation of the mean number of requests $E[c_k(\Delta T)]$ per interval $\Delta T$ may become more complex for arbitrary inter-request time distributions.

Markovian arrival processes (MAP) are included as special case with inter-request times being composed of exponentially distributed phases [12][40][51][98]. MAPs have a finite state representation with superpositions being included in the same MAP class, even if the number of Markov states is the product of the number of states of all components, leading to large and often intractable state spaces.

A remarkable result on the effect of correlated requests on caching strategies is provided by a proof that static caching due to Eq. (1) of the most popular objects is still optimal for renewal inter-request times [45][99], if their distribution fulfills conditions of an increasing hazard rate function.

Flexible Markov renewal models for TTL analysis of correlated request sequences seem to be restricted to a fixed set of $N$ objects with independent request processes in parallel over the considered time frame. Moreover, changes in the object set and changes in the popularity of an object are required to reflect realistic request pattern, as studied in the framework of shot noise models [96][119] and for requests to Wikipedia pages [56]. The rate of requests to new objects and popularity profiles over time are then used to characterize web request pattern in detail, as addressed in Section 3.3. The shot noise modeling approach is illustrated at the bottom of Figure 14 for an object $O_m$ with request rate profile $\lambda_m(t)$.

Basu et al. [6] consider a request model with different types of frequently and seldomly requested objects. They adapt TTL values dynamically in order to achieve a predefined hit ratio and study a two-level cache system for reducing the required cache storage space. Based on request traces from Akamai caches, they show that TTL adaptations can outperform LRU caching by far, where a comparison to flexible score-based methods or bounds is missing, for which large gains over LRU are reported as well [2][38][89].

*6.5. TTL Analysis Extensions for Caching Networks*

Extensions of TTL caching analysis are applied to a series of caches in a line and to feedforward networks including hierarchical, tree-shaped topologies. Responses from a set of distributed caches are evaluated. Redundant content with multiple copies of an object in different caches can be considered for resilience. Hierarchical network analysis approaches, such as [8][12][26][27][34][49][110], have to include

- superpositions of request streams as the input to caches on different hierarchy layers, and
- splitting processes for filtering the responses to be handled per cache,

from which the performance of distributed CDN, cloud and ICN architectures are estimated [6][20][23][30][64][79][130]. The optimization of distributed caching systems often leads to complex and NP-hard problems, which are approximately solved by heuristic algorithms, partly with guarantees of, e.g., an $1 - 1/e$ ratio to the optimum solution [64][79]. On the whole, the optimization of caching networks has to regard many aspects of distributed resource provisioning in a cost versus benefit trade-off, going beyond the scope of this overview of analysis approaches for single caches. Efficient content distribution in caching networks involves decisions

- on the placement of caches on nodes and/or data centers within clouds on a predefined network topology,
- on the distribution of content within installed caching infrastructures with redundancy for high content availability,
- on the cost optimization of distributed caching including request routing and retrieval,
- on the optimization of distributed caching regarding QoS aspects, such as high throughput and low delays.

*6.6. Combining Caching Strategies with TTL for Consistency*

In Sections 3-5, performance results of strategies for content selection and eviction in caches of fixed size were analysed without regard to TTL. When those strategies are combined with TTL restrictions, their hit ratio will decrease because a fraction of the hits will be invalidated due to TTL expiry.

As a basic and worst case estimation of the effect of additional TTL conditions, the hit ratio will drop by a factor of $h^{TTL,Miss}(\Delta T)$, $h^{TTL,Req}(\Delta T)$, or $h^{TTL,PR}(\Delta T)$ given in Eq. (21-26). We obtain, e.g., $h_{\text{LFU,TTL}} = h_{\text{LFU}} \cdot h^{TTL,Req}(\Delta T)$ as reduced LFU cache hit ratio due to TTL restrictions for data consistency with constant TTL intervals $\Delta T$ and TTL resets per request. The estimation is valid, if decisions on cache content are independent and unaware of TTL restrictions. On the other hand, GreedyDual, score-based, and ML methods can improve the worst case TTL impact by involving the current TTL per object in caching decisions. Then objects with expired TTL are assigned a zero score and scores are already downgraded, when the TTL comes close to expiry. TTL-aware score-based strategies will prefer objects with longer TTL as cache content to improve the hit ratio.

Nonetheless, we still almost fully agree to the statement [112] *"However, the cache consistency algorithms are not typically well integrated into cache replacement algorithms. The published work on the topic usually considers the two algorithms as separate mechanisms and studies one of the two in isolation."* Besides [112][126], the work by Berger et al. [12] is another combined approach that considers two TTL timers per object, one for the control of the amount of data in the cache and a second for data consistency.

*6.7. Combining Caching Bounds with TTL for Consistency*

Extensions for including TTLs in Belady's algorithm and knapsack bounds for clairvoyant caching are straightforward. Using Belady's algorithm or the approach [54], we can easily check if the current TTL for an object expires before its next request. In this case, the object can be evicted or marked as primary eviction candidate until it is renewed.

On the other hand, the bounds for IRM request pattern in Sections 3.1-3.2 correspond to static caching over time, without flexibility for including TTL values. The only useful correction is then to adapt the request count per object, i.e., a request is not counted if the TTL has expired beforehand.

The 2D-knapsack bounds on caching performance in Section 4 assume constant object value over time. However, knapsack bounds are fully flexible to cover score-based strategies with updates of object values and other properties per request. Therefore, the score function $S(I_{j,k}) = v_k/(\Lambda_{j,k} s_k)$ in Section 4.1 can be extended with different caching values $v_{j,k}$ per request interval $I_{j,k}$: $S^*(I_{j,k}) = v_{j,k}/(\Lambda_{j,k} s_k)$. The algorithms for computing the 2D-knapsack bounds directly include such an extension, because the main step is to sort all intervals in decreasing order of scores and then to place intervals and corresponding content into the 2D-knapsack due to the score ranking $S^*(I_{j,k})$. Different caching values per request are also useful for other purpose, e.g., for varying benefit depending on the time of day or on a current congestion status.

A different and even simpler approach for optimized caching control is proposed by Carra et al. [20]. They assume that a cache provider can order storage space of arbitrary size from huge cloud providers, where the pricing model assigns costs, which are proportional to the storage space and the duration of the order. For storage being ordered per request, the optimum clairvoyant caching strategy will store an object from the current until the next request, if the costs are lower than the caching benefit due to a hit at the next request. Then optimum content selection decision could be taken online at low $O(1)$ effort, whereas optimum solutions for strictly limited cache space involve more complex offline algorithms [20][59]. However, Carra et al. [20] admit that cache storage orders of cloud providers are performed for GByte or larger blocks and in time frames of at least one hour, which is still far from storage ordering options per request and per object.

Table 1: Properties of caching methods, performance analysis and bounds

| Methods | Data Basis for Cache Management (×): is possible as an extension | | | | | | Update Speed Compared to LRU, Complexity per Request | Replacement per Cache Miss (RpM) or Admission and Replacem. (A & R) | Approaches IRM Hit & Value Ratio Bound for Static Cache Content | Convergence Speed to IRM Steady State & for Change to a New Working Set | Hit Ratio Analysis Approaches *: for Independent Requests (IRM) |
|---|---|---|---|---|---|---|---|---|---|---|---|
| | Past Req. | Req. Count | Size | Value, Cost, Utility | TTL | Future Req. | | | | | |
| LRU | × | | | | | | $O(1)$ | RpM | No | Fastest | Markov & Approx.* |
| FIFO, Clock | × | | | | | | Faster, $O(1)$ | RpM | No | Fast | Markov & Approx.* |
| Random | × | | | | | | Slower, $O(1)$ | RpM | No | Slower | Markov & Approx.* |
| Multi-level FIFO, Random, CpR, etc. | × | | | | | | Slower, $O(1)$ | RpM | Closer to the Bound than Single Level | Multi-level Slower than Single Level | Markov* |
| LFU | | × | | | | | Slower, $O(1)$ | A & R | for Unit Size | Slow | Static (1D-)Knapsack* |
| Window-LFU | × | × | | | | | Slower, $O(1)$ | A & R | Approx. for Unit Size | Slow | |
| Tiny (W-)LFU | × | × | | | | | Slower | A & R | Approx. for Unit Size | Slow | |
| Score-gated Clock | × | × | × | × | (×) | | Similar, $O(1)$ | A & R | Approximate | Flexible | Dyn. Knapsack Bound |
| Greedy Dual | × | × | × | × | (×) | | Slower, $O(\ln M)$ | RpM | Approximate | Flexible | Dyn. Knapsack Bound |
| Utility Maximization, Hyperbolic | × | × | × | × | (×) | | Slower, $O(1)$ | A & R | Approximate | Flexible | Dyn. Knapsack Bound |
| Machine Learning | × | × | × | × | (×) | | Slower, $O(1)$ | A & R | Approximate | Flexible | Dyn. Knapsack Bound |
| Time-aware LRU | × | | × | | × | | Slower, $O(1)$ | A & R | No | Fast | Timing Analy., Markov |
| Time-to-live, TTL | | | | | × | | Depends on Timer | TTL Admission | No | Fast | Timing Analy., Markov |
| Belady's Bound | × | | | (×) | | × | Slower, $O(\ln M)$ | A & R | Beyond Static Bound | | Markov |
| Flow Opt. Bound | × | | | | | × | Slower, $O(\ln R)$ | A & R | Beyond Static Bound | | Heuristic Optimization |
| 2D-Knapsack Bound | × | × | × | × | (×) | × | Slower, $O(\ln R)$ | A & R | Beyond Static Bound | | 2D-Knapsack Heuristic |

As a final remark, there is a trade-off in web caching goals for low delay versus low traffic load, which can lead to contrary strategies for validation and expiry of cached objects. If traffic load reduction is the main goal, then data transfers within the caching architecture should be delayed until new requests for invalid data make them necessary. However, if low delay is the primary goal and transport costs are low, the data in the caches should be kept valid by forwarding the required updates as soon as possible. This improves the cache hit ratio on account of more CDN internal traffic for partly unused updates. If both, low delay and low update traffic are relevant, a cost/benefit estimate in the object scores should reflect this trade-off in caching decisions [36][57].

## 7. SUMMARY OF MEASURES FOR EFFICIENT CACHING

The hit and value ratios are in the main focus of analytic work on caching methods, whereas not many studies are also addressing the other measures of efficient caching listed in the criteria (2)-(5) of Section 2. We briefly summarize results on the update speed, overhead and flexibility of strategies for cache optimization regarding predefined goals.

In Table 1, we give an overview about the meta-data used by main caching methods and their properties with regard to the other criteria of Section 2. We refer to Berger et al. [10] for an overview table that includes over 30 caching schemes with less details on their properties. Our focus in Table 1 is more on methods that are accompanied by analytic evaluation, as indicated in the last column. Time aware LRU (TLRU) [14] is included with partial awareness of object properties as well as the class of machine learning strategies.

### 7.1. Update Speed and Upload Efficiency of Caching Methods

The requirement 2-(3) for high update speed is evaluated in the literature mainly by classifying the strategies into two categories of constant $O(1)$ versus logarithmic complexity per request. The complexity is logarithmic $O(\log(M))$ in the cache size $M$, when a sorted list of the cached objects is maintained, as required for GreedyDual methods [1][2][17][38][67] or Belady's algorithm [7][54]. On the other hand, an overview table provided by Berger et al. [10] assigns 21 out of 33 proposed caching methods to the constant $O(1)$ complexity class, fulfilling a "*low-overhead design goal*".

However, the $O(1)$ versus $O(\log(M))$ classification is crude and there are substantial differences in the update speed of methods with constant $O(1)$ effort. A recent proposal of hyperbolic caching [15] suggests comparing a new object to a set of 64 randomly chosen eviction candidates in the cache, which requires over 100-fold more update time than LRU. Table 2 indicates the update speed of FIFO, LRU and score-gated clock, which are among the fastest schemes [58]. Therefore, the mean number of requests processed per second were measured in simulation runs for each caching strategy on usual PCs. RANDOM updates are more complex than FIFO or LRU updates, as already noticed by Belady [7]: "*The strongest argument for FIFO is the fact that it is easier to step a cyclic counter than to generate a random number.*"

Table 2: Update speed of some basic caching methods with $O(1)$ complexity per request [58]

| Update speed [requests/s] | FIFO | SGC | LRU | LFU |
|---|---|---|---|---|
| Unique object sizes | $5.2 \cdot 10^7$ | $3.5 \cdot 10^7$ | $3.0 \cdot 10^7$ | $8.7 \cdot 10^6$ |
| Different object sizes | $2.0 \cdot 10^7$ | $1.5 \cdot 10^7$ | $1.3 \cdot 10^7$ | $3.7 \cdot 10^6$ |

FIFO corresponds to a clock scheme without updates for cache hits and therefore is recommended as the fastest of the basic methods [41][126]. A cyclic clock scheme is still faster than the usual doubly chained list implementation of LRU. This still holds for score-gated clock, when a simple score function is applied, e.g., $S_k = c_k/s_k$. Moreover, an ARC implementation is reported to achieve about 1/3 of the LRU update speed [63]. Moreover, similar results are reported by Li et al. [81] also in Table 2, except for a much larger gap towards LFU. Our LFU implementation follows the proposal by Sha et al. [111] with constant $O(1)$ complexity per request.

Updates for cache misses often require much more processing than for hits, involving decisions for inserting and evicting objects and data retrieval. However, LRU updates in a doubly chained list require more meta-data manipulations for hits, i.e. for moving an object from its current position to the top. For a cache miss, LRU replaces the bottom object by the requested one and steps the top counter to the bottom without reorganizing other list pointers. Consequently, the update speed of caching methods depends on the cache hit ratio, where most strategies are faster for high hit ratio except for LRU. An approximate LRU implementation proposed by Inoue [63] avoids the overhead of a double chained list.

### 7.2. Cache Update Rate and Upload Traffic

Another seldomly considered performance overhead is the demand 2-(4) of Section 2 for low web cache upload traffic. Caches for CPU support have to load every missed record to perform the next processing steps, but web caches can select and admit proper content, such that upload overhead can be reduced as an additional goal. Upload traffic and processing for large web objects causes much higher effort than loading small size records locally for CPU support. Replacement schemes like FIFO, RANDOM, LRU and GreedyDual put each missed object into the cache. Then one object is loaded per cache miss and the upload ratio equals the cache miss ratio. Score-gated [57] and LRFU [77] methods admit only high scored objects to the cache and avoid uploads of one-timers. When the score ranking of objects becomes stable over time, these content selection methods need only few uploads. A recent study on the use of hysteresis for caching by Domingues et al. [36] enables TTL-based control of the churn in web caches via reinforced counters. In this way, a minimum sojourn time for each object in the cache as well as a minimum absence time can be assigned. Then a stretch of both time periods for staying in and out of the cache by the same factor reduces the churn and the cache update/upload rate.

Figure 15 shows the cache upload ratio for different strategies as the fraction of requests with an upload. A trace with almost 500 million requests to about 5 million data objects of different size is evaluated [56]. The curves for FIFO, LRU and GreedyDual reflect the miss ratio of those methods per request. GreedyDual and Score-Gated Clock have almost the same miss ratio, but SGC admits a new object in only 1-2% of the requests.

On the other hand, score-gated caching is faster on account of approximate sorting of the objects, depending on the scores, which can be flexibly chosen for approximations of (Window-)LFU and for many cache optimization goals. The IRM hit ratio of both, (Window-)LFU and corresponding SGC approximations converge to the maximum hit ratio of Eq. (1) [57]. However, the IRM request count $c_k$ of an object $O_k$ is binomially distributed for (Window-)LFU and SGC with $p_k$ and the window size $W$ of requests from the past as parameters. This causes statistical deviations from the popularity ranking due to the IRM request probabilities $p_k$. Currently, work on the impact of approximate ranking of the hit/value ratio seems missing, i.e. to decide, if a check of a single eviction candidate is sufficient as for SGC [58][63], or if it is worthwhile to check many candidates as for hyperbolic caching [15].

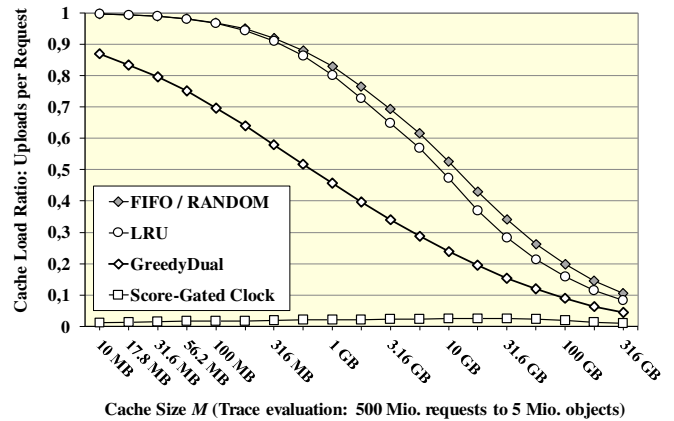

Figure 15: Upload rate per request for basic caching methods

### 7.3. Meta-Data Overhead for Cache Management

The amount of meta-data for executing a caching method should be kept as small as possible. The meta-data includes a structure for keeping cached data sorted according to a ranking principle. Score-based and ML methods maintain more detailed data about relevant properties of each object, such as size, request count, cost/benefit estimates. The RANDOM strategy needs a minimum of meta-data only for checking if a request can be served from the cache.

Such meta-data is at least available for $M$ objects in the cache, or may be stored for all objects that were requested in the past. Therefore, the storage space complexity is in the range $O(M)$ - $O(N)$. Strategies with meta-data only for cached objects delete information about evicted objects, which seem to be new when they reappear at their next requests. As a compromise, meta-data should cover a relevant working set, which can be limited to, e.g., $3M$ objects to stay within $O(M)$ meta-data overhead. Some strategies require additional data, such as the Window-LFU strategy, which has to store object identifiers for a window of $W$ past requests.

While meta-data overhead is crucial on local devices with limited storage, it is less relevant for web cache installations in large and highly performant data centers. Compared to the typical size of web data objects in a range of kB-MB as reported in many studies [10][112][116][130], a few additional Bytes per object for cache management seem negligible.

A TinyLFU approach [37] proposes data compression and reduction schemes via Bloom filter and other measures. However, then the (de-)compression effort can slow down the cache update speed. Moreover, imprecise or missing information in compressed format can lead to suboptimal content selection, which affects hit and value ratios. Thus, the approach is useful only under stringent storage limitations, e.g. on low power devices or processors with fast but small storage units.

### 7.4. Modeling of Dynamics in the Object Popularity and Correlated Request Sequences

Regarding the demand 2-(2) of Section 2, we notice a lack of a clear picture for modeling the dynamics and correlation in web request pattern. Even if dynamics in the popularity of web objects is moderate as reported in measurement studies of web request traces [2][56][89][119], a web cache strategy has to respond to recent trends. (Semi-)Markov processes can model correlated request intervals [6][8][12][45][61]. The Markov results of Section 5 based on independent requests are extensible to $K$-state Markov request processes [107]. Their combination with Markov analysis of FIFO, CpR, RANDOM or LRU cache content would extend the IRM approaches to $K$-fold state space [61]. However, whether and how to extend the product form solutions to Markov representations of correlated requests is a topic for future study.

Shot noise models [96][119] are another approach to describe correlated request pattern aggregated from individual content request streams, as discussed in Sections 3.3 and 6.4, where a tractable analysis depends on simplifying assumptions about the request rate profiles.

For the request pattern of most cache applications, request traces seem only sporadically available to verify models and parameter sets for measuring the effect of request dynamics on the hit and value ratio. Consequently, providing a solid basis of benchmarks as learning sequences for popularity prediction remains an open challenge, as pointed out e.g. in the conclusions of Section 3.4 of [101]. High temporal request dynamics can reduce the hit ratio and impede predictions. Optimum caching strategies in adaptation to the specific request dynamics are another future research topic.

Finally, Table 3 gives an overview of the modeling and analysis approaches addressed in Section 3 - 6 with regard to

- the underlying request pattern (IRM, correlated, general),
- the flexibility to include Byte and value hit ratios,
- the computational complexity, and
- the scope of the results, i.e. for which caching methods.

Table 3: Scope and complexity of the presented cache modeling and analysis results

| Analysis Results & Approximations | Reference for Detailed Results | Scope: General Request Pattern (GRP) or IRM | Fixed or Variable Object Size OHR, BHR, VHR: Object/Byte/Value Hit Ratio | Computational Complexity ($N, R$: Number of Objects, Requests) ($M$: Cache Size; $M^*$: Mean Number of Objects in the Cache) | Applicability for Caching Methods |
|---|---|---|---|---|---|
| Knapsack Bound for Static Caching | Sect. 3.2, Eq. (5) [4][116] | IRM | Variable Object Size OHR, BHR, VHR | $O(N \log N)$ | General, Non-Predictive |
| Knapsack Bound for Varying Object Popularity | Sect. 3.3, Eq. (7) | Popularity $p_k(R)$ of $O_k$ is Varying per Request $R$ | Variable Object Size OHR, BHR, VHR | $O(R \cdot N \log N)$ | General, Non-Predictive |
| Belady's Bound | Sect. 4.3 [7][86][90] | GRP | Fixed Size, OHR | $O(R \log M)$ | General, Clairvoyant |
| 2D-Knapsack Bound | Sect. 4.1-4.2 [59] | General Request Pattern | Var. Size, OHR, BHR, VHR | $O(R \cdot (M^* + \log R))$ | General, Clairvoyant |
| Min. Cost Flow Bound | [11] | General Request Pattern | Variable Object Size, OHR | $O(R \log R)$ | General, Clairvoyant |
| Product Form Solution for FIFO, RANDOM, CpR | Sect. 5.1, Eq. (11) [52][72][73] | IRM | Fixed Size OHR | $O(M \cdot N)$ | FIFO, RANDOM, CpR & Variants |
| Product Form Solution for Caches with $K$ Levels/Segments | Sect. 5.6, Eq. (20) [50][78] | IRM | Fixed Size OHR | $O(N \cdot K^2 (M/K)^K)$ | $K$-Level FIFO, CpR, RANDOM Caches |
| Analytic LRU Solution Including Cache Filling Phases | Sect. 5.2-5.3, Eqs. (12-17), [72] | IRM | Variable Object Size OHR, BHR, VHR | $O(N!/(N-M)!)$ | LRU |
| Approximations by Fagin, Che, Dan & Towsley | Sect. 6.3 [23][32][42] | IRM | Variable Object Size OHR, BHR, VHR | $O(N)$ | LRU, FIFO, CpR, RANDOM Caches |
| TTL Hit Ratios for Poisson Request Process | Sect. 6.2 [26][27][34] | IRM | Variable Object Size OHR, BHR, VHR | $O(N)$ | TTL Caching |

### CONCLUSIONS

We have addressed the analysis of caching performance from initial to advanced recent work in three main fields:

- General knapsack and minimum cost flow bounds on the cache hit ratio for optimum caching value or utility,
- Markov analysis results for a set of basic caching strategies with extensions to multi-level/-segment caches, and
- Time-to-live caching with variants for adaptive cache load control and for enforcing data consistency.

Knapsack solutions provide upper bounds on caching performance in a simple format for static cache content, when requests are independent, and in a 2-dimensional format for arbitrary request sequences. The latter approach leads to a general hit/value ratio bound for dynamic caching strategies, including Belady's algorithm as special case for unit data size.

A broad class of utility- and score-based policies, GreedyDual and machine learning methods enables content selection for optimized web caching efficiency with regard to network-wide goals on costs, benefit, QoS and/or traffic load. Although basic LRU, FIFO, clock and TTL strategies without awareness of the size, value, request count etc. per data object are often subject to severe performance gaps for specific web caching goals, they are still widely used because of their simple implementation and low update effort.

For performance evaluation, a common product form solution is valid for FIFO, clock-based and RANDOM policies via Markov analysis in steady state for independent requests. The solution is scalable for large caches, and extends to multi-level/-segment caches. Another Markovian steady state hit ratio formula for LRU extends to cache startup phases and to caches with objects of varying size, but is tractable only for small caches. LRU hit ratio approximations perfectly complement the Markov analysis results, because the accuracy improves towards asymptotic exactness for large caches.

The analysis of time-to-live caches starts from a simpler concept for controlling each data object separately by a timer, when sufficient cache storage is available. TTL caches directly apply for the validation of data, and for DNS and other web applications with frequently updated data of small size. Adaptive TTL-based cache load control and many extensions of TTL concepts are considered in a broadening TTL cache analysis workstream. This includes hierarchical caching networks and approximations for LRU and FIFO caching performance. TTL conditions can be easily included in general knapsack bounds. On the other hand, studies for combining TTLs for data consistency with TTL load control are still rare as well as for the analysis of basic strategies for fixed cache size with TTL restrictions for consistency on top.

While the hit ratio of basic methods is in the main focus of most caching analysis approaches, other aspects need more attention, such as the update speed, network-wide resource optimization and end-to-end service quality, which require advanced and flexible caching strategies. Correlated request pattern beyond IRM are partly addressed by Markov, shot noise and other models, but their conclusions for optimized caching strategies seem complex with a need for clearer characterization of main impacts as a future research topic.